\documentclass{jfm}
\usepackage{url} 
\usepackage{graphicx}
\usepackage{epstopdf}
\usepackage{subfig,float}
\usepackage{amsmath,amssymb}
\usepackage{natbib}
\usepackage{xcolor}
\usepackage{dcolumn}%
\usepackage{bm}%
\usepackage{cleveref}
\crefname{section}{§}{§§}
\Crefname{section}{§}{§§}\newcommand{\lra}[1]{\langle #1 \rangle }

\newcommand{\bds}[1]{\boldsymbol{#1}}

\newcommand{\fltr}{\overline}
\newcommand{\grad}{{\mbox{\boldmath $\nabla$}}}

\newcommand{\fu}{\fltr\bu}

\newcommand{\btau}{{\mbox{\boldmath $\tau$}}}
\newcommand{\br}{{\bf r}}

\newcommand{\bu}{{\bf u}}

\newcommand{\bx}{{\bf x}}

\def\be{\begin{equation}}
\def\ee{\end{equation}}

\begin{document}

%\preprint{}
\shorttitle{Turbulent bubbly flows}
\shortauthor{A. Innocenti, A. Jaccod, S. Popinet, S. Chibbaro}

\title{Direct Numerical Simulation of bubble-induced turbulence} %Title of paper



\author{Alessio Innocenti \aff{1}, Alice Jaccod\aff{1}, St\'ephane Popinet \aff{1}
\and Sergio~Chibbaro\aff{1}
}
\affiliation{\aff{1} Sorbonne Universit\'e, CNRS, UMR 7190, Institut Jean Le Rond d'Alembert, F-75005 Paris, France}
%\thanks{}


%\date{\today}

%\pacs{}

\maketitle

%\usepackage{geometry}
% \geometry{
% a4paper,
% total={170mm,257mm},
% left=20mm,
% top=10mm,
% right=20mm,
% }
%
%\linespread{1.5}

\begin{abstract}
We report on a investigation of turbulent bubbly flows.
Bubbles of a size larger than the dissipative scale, cannot be treated as point-wise inclusions, and generate important hydrodynamic fields in the carrier fluid when in motion.
%In liquid-gas multiphase flows, buoyancy is the typical force which makes bubbles rise in the surrounding liquid.
Furthermore, when the volume fraction of bubbles is large enough,  the bubble motion may induce a collective agitation due to hydrodynamic interactions which display some turbulent-like features.
We tackle this complex phenomenon numerically, performing direct numerical simulations (DNS) with a Volume-of-fluid (VOF) method. %Adaptive grids are used to make the computational effort feasible when needed. 

In the first part of the work, we perform both 2D and 3D tests in order to determine appropriate numerical and physical parameters.
%Specifically we have studied the rise of an array of bubbles. 
%Particular attention has been paid to the effect of grid resolution and the value of the physical parameters, notably the density and viscosity ratio between the two phases.
These tests confirm that realistic results can only be obtained if realistic physical parameters are set and appropriately resolved by the numerics. 
These resolution constraints are increasingly severe with increasing velocity of the bubbles. 

We then carry out a highly-resolved simulation of a 3D bubble column, with a configuration and physical parameters similar to those used in laboratory experiments.
This is the largest simulation attempted for such a configuration and is possible only thanks to adaptive grid refinement. 
Results are compared both with experiments and previous coarse-mesh numerical simulations. 
In particular, the one-point Probability Density Function (PDF) of the liquid velocity fluctuations is in good quantitative agreement with experiments, notably in the vertical direction, although more extreme events are  sampled in the present configuration.
The spectra of the liquid kinetic energy show a clear $k^{-3}$ scaling.
%Bubbles initially placed at the bottom of the column rise freely afterwards.
%, that is the Reynolds number based on bubble diameter is high.
%We have analysed both the spectra and the statistics of fluid velocity. 
%At smaller scales viscous dissipation seems to dominate.
%as a function of the wavenumber.
The mechanisms underlying the energy transfer and notably the possible presence of a cascade are unveiled by a local scale-by-scale analysis in the physical space.
The analysis shows that the spectra are related to a transfer of energy between scales, for which there is a range where the turbulent flux and the dissipation  are in balance.
In our unsteady configuration it is found that the flux may take both positive and negative values.
The comparison with previous simulations indicate to what extent simulations not fully resolved may yet give correct results, from a statistical point of view. 
\end{abstract}

\section{Introduction}

Multiphase flows are common and a central topic in fluid mechanics~~\citep{prosperetti2009computational}. They are present in a number of phenomena including  pollutant dispersion, sedimentation, bubble spray in ocean dynamics, and bubble columns.

Among the various kinds of multiphase flows, bubbly flows are a particularly challenging and key field of investigation, both for their fundamental dynamics and their numerous applications in engineering and environmental science~~\citep{Prosperetti:2004p5898,clift2005bubbles,magnaudet2000motion,ern2012wake,lohse2018bubble}. 
While much attention has been paid in the last decades on the dynamics of small inertial or neutrally-buoyant particles ~\citep{crowe1996numerical,balachandar2010turbulent,maxey2017simulation,elghobashi2019direct}, much less is known for bubbles, because they are experimentally, numerically and theoretically more complex ~\citep{prosperetti2017vapor,mathai2020bubble}.

Bubbles have a deep impact on the characteristics of the flow.
When injected at the bottom of a liquid-filled
vessel, the bubbles rise due to buoyancy. 
High transfer rates can then be attained owing to the increased contact area between the gas and liquid phases,
and to the liquid agitation induced by bubble motion. These efficient heat and mass transfer characteristics make bubble columns useful as industrial devices. 
%They are also a good configuration to analyse the fundamental features of bubble dynamics, as highlighted in the pioneering work of ~~\cite{lance1991turbulence}.
Since a reliable prediction
of bubble residence time and available interfacial area is crucial for an accurate design of industrial devices, the understanding of bubble flow dynamics is essential.
In particular, due to the large density difference between gases and liquids, under the effect of buoyancy the bubbles rise with a velocity considerably different from that of the liquid. 
They thus induce velocity disturbances in the liquid that collectively generate a complex agitation,
referred to as bubble-induced agitation or pseudo-turbulence and this effect
 is key to grasp the physics of more complex flows. 
We focus in this work on this phenomenon, leaving out for the moment the presence of other effects such as the surrounding background turbulence.

Several experimental studies have been carried out to investigate this particular regime in different configurations~\citep{zenit2001measurements,martinez2007measurement,riboux2010experimental,mendez2013power}, and significant progress has been made in figuring out the characteristic features of bubble-induced agitation ~\citep{risso2018agitation}.
In particular, it is an experimental evidence~\citep{risso2002velocity} that at moderate-to-large Reynolds numbers ($Re\gtrsim 100$) the wakes of interacting bubbles are attenuated, which tends to show that at large Reynolds numbers the dominant mechanism underlying liquid agitation is 
nonlinear wake interactions.

Focusing on the liquid fluctuations induced by the bubbles, 
the key observations are that (i) the probability density function (PDF) of the vertical fluctuations is strongly skewed and thus non-Gaussian, while the horizontal one is symmetric; (ii)
The energy spectrum of the liquid agitation $E(k)$ displays  a robust scaling $E\sim k^{-3}$.
Some issues remain unclear however.
The range where this scaling applies is under discussion, with some experiments pointing to larger scales than the diameter~\citep{riboux2010experimental}, while others found the scaling at smaller scales~\citep{mercado2010bubble,prakash2016energy}. Moreover, in some experiments a Kolmogorov spectrum 
$E\sim k^{-5/3}$ is observed either at small~\citep{riboux2010experimental} or large scales~\citep{lance1991turbulence,prakash2016energy}.
From a physical point of view,  two main mechanisms appear to underly these scalings: the superposition of Gaussian fluctuations generated near the bubbles ~\citep{risso2011theoretical}, and the turbulent fragmentation ~\citep{lance1991turbulence} notably at high Reynolds. It is difficult to disentangle these two mechanisms, and the steep spectrum $E\sim k^{-3}$ corresponds to a smooth flow ~\citep{Mon_75}, which could be related to a number of different situations ~\citep{boffetta2012two}.

Fluid turbulence can be characterised by a \emph{cascade} phenomenon, which is first of all related to a constant flux of kinetic energy towards a certain range of scales ~\citep{Fri_95,boffetta2012two,alexakis2018cascades}. The existence of such a cascade in the pseudo-turbulence regime characteristic of bubble columns would help to understand the underlying mechanisms. Because the injection of energy is made via buoyancy, it is not clear \emph{a priori} which scales are forced and toward which scales the energy is transferred.
The following issues thus remain to be addressed: (a) Is there a cascade of energy in the sense of a non-zero flux triggered by bubbles? (b) What is the direction of the cascade? (c) How is the phenomenology of the spectrum of velocity fluctuations related to this picture? (d) What are the mechanisms underlying the modulation of the flow by bubbles? (e) What is the influence of Reynolds number?

The purpose of this work is to address these issues with unprecedented high-resolution numerical simulations, combining several 2D and 3D numerical experiments.
Indeed numerical simulations allow to access the dissipation and energy fluxes which are difficult to measure experimentally.

Turbulent bubbly flows display a strong multi-scale character with a very broad spectrum of scales, including the excited fluid modes~\citep{Pope_turbulent} and the scales related to bubble boundary layers~\citep{tryggvasondirect}. 
In addition the density ratio between the two phases is generally very high (about $1000$ in experimental flows) which makes the problem stiff.
These contraints make direct numerical simulations of turbulent bubbly flows particularly challenging.

Recent attempts to address similar issues through numerical simulations include ~\citet{roghair2011energy} and \citet{pandey2020liquid}. 
In all these studies the resolution has been kept at about 20 points per diameter ($\Delta x=d_b/20$), independently from the Reynolds number of the bubbles which is larger than 200 in most of the cases. \citet{cano2016paths} have shown however that such a resolution does not allow to properly resolve the boundary layers around bubbles at high Reynolds number and this may lead to quantitative and even qualitative errors on the dynamics. The resolution should rather increase proportionaly to the bubble Reynolds number.

Furthermore in the first work ~\citep{roghair2011energy} realistic physical properties are chosen but just a few bubbles are released, of the order of $10$, while in the second study many bubbles are followed but with a very low density ratio between the fluid and the gas, between 1.1 and 20. 
The nonlinear interactions among bubbles are however key to the dynamics and their statistical study requires the presence of a large number of bubbles~\citep{risso2018agitation}. 
Moreover, while in some cases and with respect to specific observables the correct physics 
 may be reproduced with a low density ratio~\citep{Diotallevi:2009p1553}, that cannot be claimed in general and requires further scrutiny. 

 As a matter of fact, these numerical simulations are implicitly coarse-grained, and therefore they should be considered as Large Eddy Simulations (LES) rather than DNS.
 As for LES of single-phase flows, results may well be in accordance with experiments but comparison with resolved DNS appears necessary~\citep{Pope_turbulent}.
A  purpose of the present work is precisely to provide a reference high-resolution DNS and to compare with previous works to assess to which extent coarse simulations are reliable.
%While it has been already found that single bubble trajectories are not correctly reproduced with coarse-grained simulations ~\citep{cano2016paths}, the behaviour of collective statistical observables might be more robust.
% It would be highly relevant if it turns out that correct statistics may be found in low-resolution numerical simulations.  

From a numerical point of view, different techniques can be used to study interfacial flows~\citep{tryggvasondirect,popinet2018numerical}. In the present work, we use the Volume-Of-Fluid (VOF) open-source code Basilisk~\footnote{\url{http://basilisk.fr}}, which provides efficient adaptive mesh refinement, a key requirement to perform the high-resolution 3D bubble column simulations presented here.

The detailed contents of this paper are the following:
In \cref{sec:mat} we review the basic mathematical framework of the problem, with particular attention to the different non-dimensional parameters relevant for the physics of bubbly flows.
In \cref{sec:num}, we briefly introduce the Basilisk code, giving some information about the grid-refinement method. 
The numerical schemes used for the integration of the equations are given together with the main references.
In \cref{sec:test}, we present the results obtained in a series of 3D tests at low or moderate Reynolds numbers.  These tests consist in a regular array of rising bubbles, and we compare our results against analytical predictions in the case of Stokes flows, or to previous numerical studies.
These tests are important not only to assess the different numerical codes but also to analyse the interplay between the physical parameters, namely Reynolds, density and viscosity ratios and the numerical requirements to get accurate results.
In \cref{sec:2D} we show the results obtained with very high-resolution simulations of a 2D bubble column at different Reynolds numbers.
Since with the the present computational capability, it is not possible to carry out a parametric analysis of a realistic flow in 3D, these simulations have been used 
to accurately set the numerical and physical parameters to be used in a single 3D simulation. 

We show both unsteady and steady simulations to verify that a reasonable  convergence in the relevant statistics is obtained also in the unsteady cases.
Different statistical observables are studied, namely spectra of kinetic energy at different Reynolds, and the one-point PDF of the velocity both in the horizontal and vertical direction.

The 3D bubble-column case results are reported in \cref{sec:3D}.
The configuration corresponds to an Archimedes number of $Ar=185$ and is globally comparable with typical laboratory experiments.
The PDF of the velocity is analysed first and compared with previous experimental results~\citep{riboux2010experimental}.
The spectrum  of the kinetic energy is then computed and compared with results obtained very recently at low resolution by \citet{pandey2020liquid}.
To gain physical insights and address the issues related to the \emph{cascade}, we present a scale-by-scale analysis of  the energy transfers in physical space, rather than in spectral space as commonly done in isotropic turbulence.
This multi-scale approach has been developed  in relation to the filtering used in LES   ~\citep{germano1992turbulence}. 
This approach allows to simultaneously measure the effect of the small filtered scales on the unfiltered scales, and to correlate it with different observables of the large scales, which permits the detailed study of the cascade process in different situations ~\citep{borue1998local,meneveau2000scale,chen2003joint, Chen:2006p1741, Eyink:2006p1379, chen2006kelvin,eyink2006multi,alexakis2018cascades}.
Furthermore, in contrast with the spectral approach, this method is by definition local in space and is thus not limited to homogeneous flows ~\citep{eyink2009localness,aluie2009localness}.
A conclusion \cref{sec:conc} summarizes and discusses our findings.

Three appendices provide some  complements for the results shown in the main text; some more comparison with the literature is given for the case of the array of bubbles (Appendix \ref{app:array}); some numerical issues, such as the effect of grid refinement are presented in Appendix \ref{app:tech}, and some complementary results for the 2D simulations are given in Appendix \ref{app:C}.

\section{Mathematical formulation}
\label{sec:mat}
\subsection{Problem statement}

We investigate the dynamics of a monodisperse suspension of bubbles rising under the action of buoyancy in a fluid initially at rest.
The density, the viscosity and the surface tension of each fluid are considered constant during each numerical experiment. 

Different configurations of increasing difficulty are studied. We first consider 
an infinite homogeneous suspension, which is represented by the periodic repetition of cubic unit cells containing a given number of bubbles.
When a single bubble is considered, this describes an ordered array of bubbles.
We then consider a more realistic configuration in which a given number of bubbles 
are initially randomly distributed at the bottom of a channel and rise through it.
This problem is analyzed in two dimensions while varying several parameters and is then studied in three dimensions at high Reynolds numbers.

Several physical parameters characterize the problem: the gas volume fraction $\phi$, the number of bubbles $N_b$, the diameter of the bubbles $d_b$ calculated as the diameter of the sphere of equivalent volume, the gravity acceleration $g$, the viscosity of the two fluids $\mu_b,~\mu_l$, their densities $\rho_b,~\rho_l$, and the surface tension $\sigma$. We use the subscripts $b$ for bubbles and $l$ for liquid.
Four dimensionless groups can be formed in addition to the number of bubbles and the volume fraction.
Two are the density and viscosity ratio $ \frac{\rho_b}{\rho_l}$ and $\frac{\mu_b}{\mu_l}$.
We briefly analyse the impact of the density ratio but in almost all simulations we have fixed ${\rho_b}/{\rho_l}=10^{-3}$ and ${\mu_b}/{\mu_l}=10^{-2}$, which are typical values for air bubbles in water.

The other two dimensionless groups can be characterised by the Galileo number
\begin{equation}
Ga\equiv\frac{\rho_l\vert \Delta \rho\vert g d_b^3}{\mu_l^2},
\end{equation}
where $\Delta \rho= \rho_b-\rho_l$, or equivalently the
Archimedes number $Ar\equiv \sqrt{Ga} $,
and the Bond (or E\"otv\"os ) number
\begin{equation}
Bo\equiv\frac{\vert \Delta \rho\vert g d_b^2}{\sigma}
\end{equation}
These numbers indicate the relative importance of buoyancy and surface tension.

When bubbles move, a typical velocity scale can be added to the 
relevant quantities.
We compute the evolution with time of the (space-)averaged velocity of each phase. We note the average bubble velocity $\lra{U_b}$. It is then possible to define the bubble Reynolds number based on this velocity
\begin{equation}
Re\equiv\frac{\lra{U_b} d_b}{\nu_l},
\end{equation}
where $\nu_f$ is the kinematic viscosity of the liquid.
It is also possible to use another group which compares inertial effects with surface tension, the Weber number
\begin{equation}
We \equiv \frac{\rho_b \lra{U_b}^2 d_b}{\sigma} = \frac{Bo Re^2}{Ga}.
\end{equation}
It is important to note that the average bubble velocity may or may not reach a stationary state in our numerical experiments, so that in general the dynamic dimensionless numbers are dependent on time $Re=Re(t)$.

As anticipated in the introduction, in the first part of this work, we analyse the relation $U=U(N_b,\phi, Bo, Ga)$
and compare the results with recent numerical studies obtained with different methods, and with analytical solutions whenever possible.
In the second part, we will study the phenomenology of some realistic configurations with the parameters set as in typical experiments.

%\subsection{Flow regimes}
%\textit{? si scrive ?}

\subsection{Governing equations}

Both fluids are governed by the Navier--Stokes equations, which we take here in the incompressible limit
\begin{eqnarray}
\nabla \cdot \mathbf{u_i} &=& 0 \\
\frac{\partial {\bf u}_i}{\partial t} + \nabla \cdot  ({\bf u}_i \otimes {\bf u}_i) &=&  \frac{1}{\rho_i} \left( - \nabla p_i + \nabla \cdot (2 \mu_i \bds{ D}_i)\right) +  {\bf f_i}~,
\end{eqnarray}
where $\bds{ D}_i= [\nabla u_i +  (\nabla u_i)^{\textbf{T}}]/2$ is the symmetric deformation tensor, the subscript $i$ indicates each phase $i={b,l}$, and $f_i$ represents the acceleration due to volume forces, which in the present case are the gravity, $\mathbf{f}_i={\bf g}$.

In addition, the appropriate boundary conditions at the interface between the phases must be imposed. Since we do not consider any phase change, the interfacial condition for viscous fluids is simply $\mathbf{u}_b = \mathbf{u}_l$, or
\begin{equation}
[\mathbf{u}]_S=0,
\end{equation}
where we have used the jump notation, \emph{i.e.} the notation $[x]_S = x_b-x_l$.
At equilibrium $\mathbf{u}=0$, the jump at the interface is given by
\begin{equation}
[p]=\sigma \kappa \mathbf{n},
\end{equation}
where $\mathbf{n}$ is the unit normal vector defined as directed outward from the bubbles, and $\kappa$ is the mean curvature of the interface. 
This set of equations are solved with the Basilisk library~\footnote{\url{http://basilisk.fr}} with the numerical methods described in the following section.
%% In particular, the accurate numerical calculation of the surface tension is key in multiphase flows ~\citep{popinet2018numerical}.

\section{Numerical method}
\label{sec:num}

Basilisk is a library of solvers written using an extension of the C programming language, called Basilisk C, adapted for discretization schemes on Cartesian grids (see \url{http://basilisk.fr}). 
Space is discretized using a Cartesian (multi-level or tree-based) grid where the variables are located at the center of each control volume (a square in 2-D, a cube in 3-D) and at the center of each control surface.
The possibility to adapt the grid dynamically is key to efficiently simulate multiphase flows ~\citep{popinet2009accurate}.
Two primary criteria are used to decide where to refine the mesh. They are based on a wavelet-decomposition of the velocity and volume fraction fields respectively ~\citep{van2018towards}. The velocity criterion is mostly sensitive to the second-derivative of the velocity field and guarantees refinement in developing boundary layers and wakes. The volume fraction criterion is sensitive to the curvature of the interface and guarantees the accurate description of the shape of bubbles. Both criteria are usually combined with a maximum allowed level of refinement. As demonstrated in previous work, using the earlier code Gerris ~\citep{cano2016paths}, this strategy leads to very large savings in computational cost compared to fixed Cartesian grid approaches.

The numerical scheme implemented in Basilisk is very close to that used in Gerris as described in ~\cite{popinet2009accurate} and many other articles. The Navier--Stokes equations are integrated by a projection method ~\citep{chorin1969convergence}.
Standard second-order numerical schemes for the spatial gradients are used ~\citep{popinet2003gerris,popinet2009accurate,lagree2011granular}.
In particular, the velocity advection term $\partial_j (u_j u_i )^{n+1/2}$ is
estimated by means of the Bell-Colella-Glaz second/third-order unsplit upwind scheme ~\citep{popinet2003gerris}. 
In this way, the problem is reduced to the solution of a 3D Helmholtz--Poisson problem for each primitive variable and a Poisson problem for the pressure correction terms. 
Both the Helmholtz--Poisson and Poisson problems are solved using an efficient multilevel solver ~\citep{popinet2003gerris,popinet2015quadtree}. 

Time is advanced using a second-order fractional-step method with a staggered  discretization in time of the velocity and scalar fields ~\citep{popinet2009accurate}: one supposes the velocity field to be known at time $n$ and the scalar fields (pressure, temperature, density) to be known at time $n-1/2$, and one computes velocity at time $n+1$ and scalars at time $n+1/2$.

The interface between the fluids is tracked with a geometric Volume-Of-Fluid method ~\citep{hirt1981volume,scardovelli1999direct}. The surface tension term is computed using an accurate well-balanced, height-function method ~\citep{popinet2018numerical}.

Periodic, no-slip and free-slip boundary conditions will be imposed in the different computations considered.

\section{Preliminary tests}
\label{sec:test}
To assess the accuracy of the numerical code for the simulation of two-phase bubbly flows, we have reproduced several literature test cases~~\citep{esmaeeli1998direct,esmaeeli1999direct,sangani1987sedimentation} which have been also considered recently with another numerical approach ~\citep{loisy2017buoyancy}. 
In particular we have focused on the configuration of arrays of rising bubbles, which, in the presence of gravity, start to rise inside a heavier fluid at rest, due to buoyancy. After an initial transient where bubbles accelerate, they eventually reach a quasi-steady-state regime. Depending on bubble size, surface tension and density, they may follow non-rectilinear paths, with periodic or chaotic lateral oscillations ~\citep{cano2016paths}. 
All benchmark tests consist in a regular array of bubbles, which is reproduced numerically using a single bubble in a periodic cell. Changing the cell size with respect to the bubble size, we can adjust the volume fraction of the array. Note that since the computational domain is unbounded in all directions, an additional body force $- \langle \rho \rangle g$ must be added to avoid that the system accelerates in the vertical downward direction.
We present briefly here only the most significant results, while more details are given in Appendix \ref{app:array}.

We have first compared our simulations with the theory of ~\cite{sangani1987sedimentation} for the Stokes flow regime. 
The configuration consists in a cubic array of spherical bubbles at different volume fractions. 
The non-dimensional numbers of the simulation are the same as in previous DNS studies, namely:
\begin{equation*}
Ar = 0.15 \qquad Bo = 0.38 \qquad \rho_b / \rho_l = 0.005 \qquad \mu_b / \mu_l = 0.01 \, .
\end{equation*}
Although at very low-Reynolds number, this is a severe test case since it is 3D and the number of points required may increase rapidly when varying the concentration. The Basilisk grid refinement algorithm is greatly beneficial in this case.
In Table \ref{tab:phi} we show the steady-state velocity of the bubble array normalized with the velocity of a single isolated bubble, and the quantitative numerical error.
The agreement between the numerical and the analytical solution $\frac{U}{U_0} = 1 - 1.1734 \mu^* \phi^{1/3} + O(\phi)$, is satisfactory.
\begin{table}
\begin{center}
\begin{tabular}{ c | c | c | c | c}
\hline
$\phi^{1/3}$ & $U/U_0$ DNS & $U/U_0$ Analytical & relative error & $d_b/Delta$  \\
\hline
$0.2$ & $0.768$ & $0.755$ & 1.7\% &$63.5$  \\
$0.3$ & $0.651$ & $0.632$ & 2.5\% & $47.64$  \\
$0.4$ & $0.525$ & $0.51$ & 3 \% & $63.5$  \\
$0.5$ & $0.408$ & $0.39$ & 4.3\% & $79.4$  \\
%\textbackslash \\
\hline
%%\bottomrule
\end{tabular}
\caption{DNS and analytical results for the test analysed by ~\cite{sangani1987sedimentation}.}
\label{tab:phi}
\end{center}
\end{table}

\begin{figure}
\center
{\includegraphics[width=.45\textwidth]{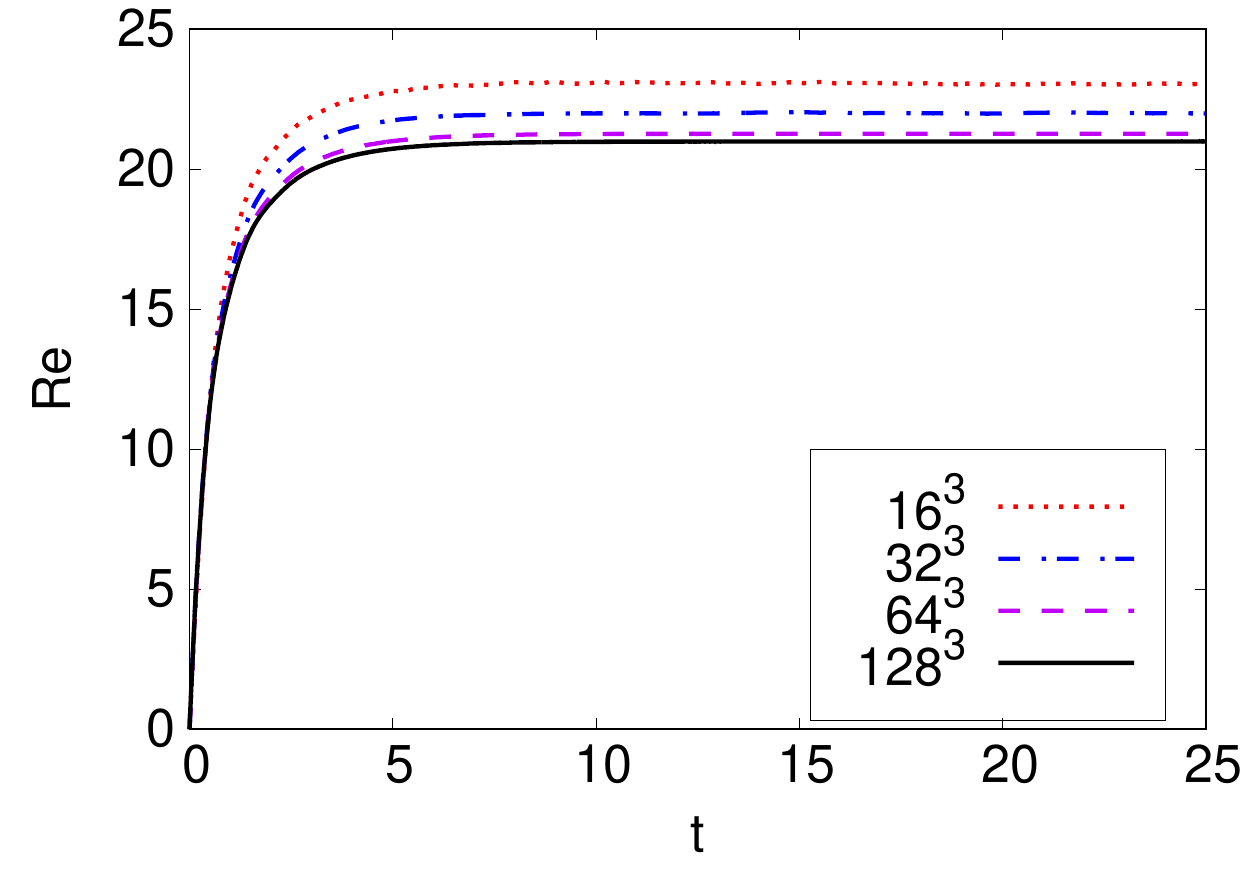}}  \hspace{1cm}
{\includegraphics[width=.42\textwidth]{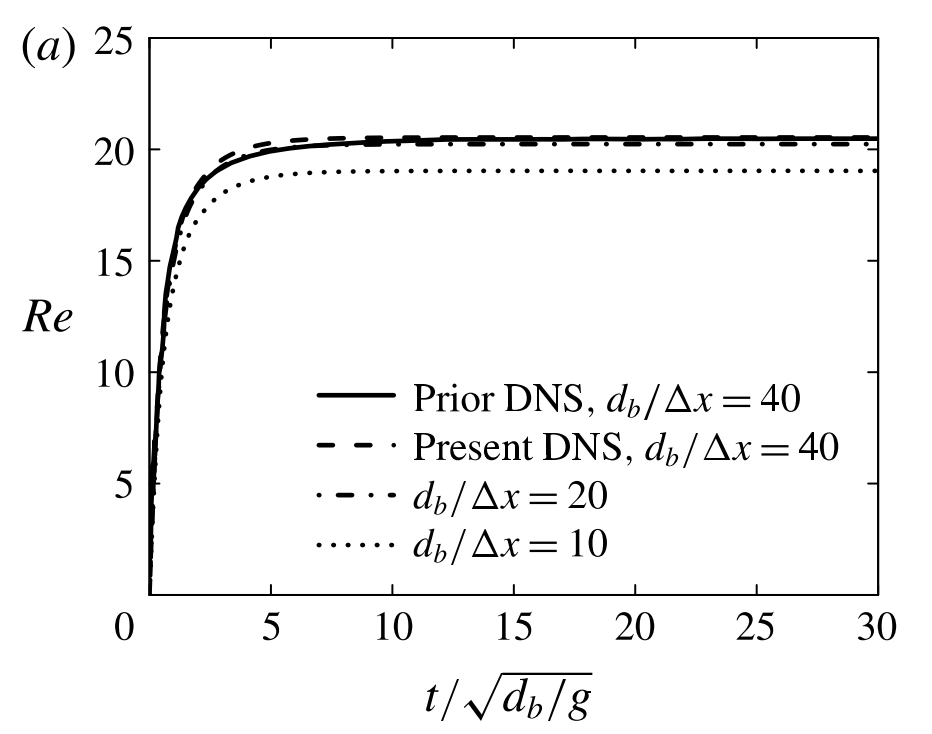}} \\
\caption{Time evolution of the bubble Reynolds number for different grid resolutions, for the 3D configuration. Present results on the left panel, Loisy \etal ~results on the right panel. In the right panel "prior DNS" stands for Esmaeeli \etal, while present DNS for Loisy \etal.}
\label{Fig:3D-tryg}
\end{figure}

We then considered test cases at finite Reynolds numbers. 
In figure \ref{Fig:3D-tryg}, results are shown for the test case proposed by ~\cite{esmaeeli1999direct}, which is 3D at moderate Reynolds numbers.
For this case the flow parameters are
\begin{equation*}
Ar = 29.9 \qquad Bo = 2 \qquad \rho_b / \rho_l = 0.1 \qquad \mu_b / \mu_l = 0.1 \, .
\end{equation*}
Our simulations are compared against both the original DNS of ~\citet{esmaeeli1999direct} and the more recent results of ~\citet{loisy2017buoyancy}. We have used a more refined grid with respect to both other DNS to test the convergence in a rigorous way. 
Results are in good agreement, but grid convergence is not achieved with the number of points given in previous works ~\citep{esmaeeli1999direct,loisy2017buoyancy}, where the authors indicate that $30$ points per diameter are sufficient.

\begin{figure}
\center
{\includegraphics[width=.45\textwidth]{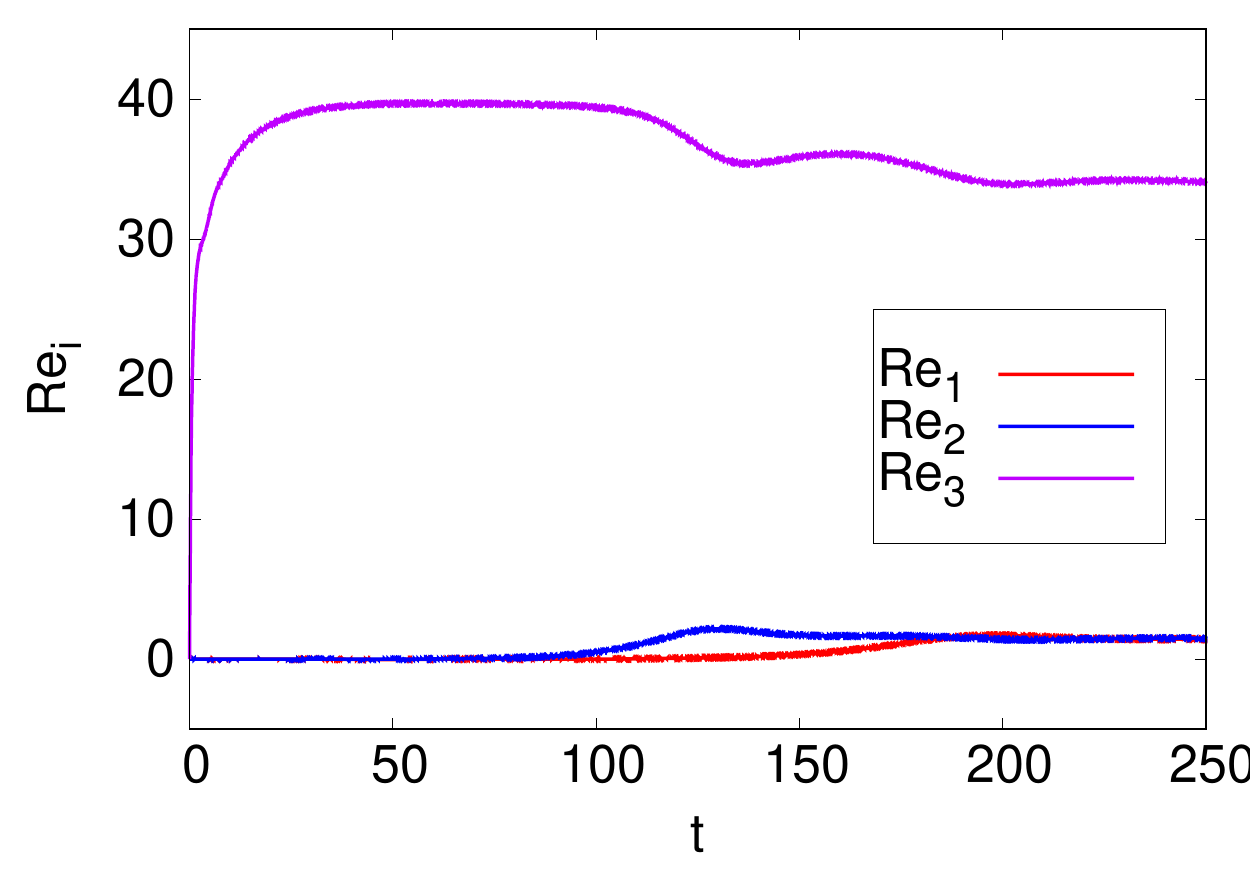}}  \hspace{1cm}
{\includegraphics[width=.45\textwidth]{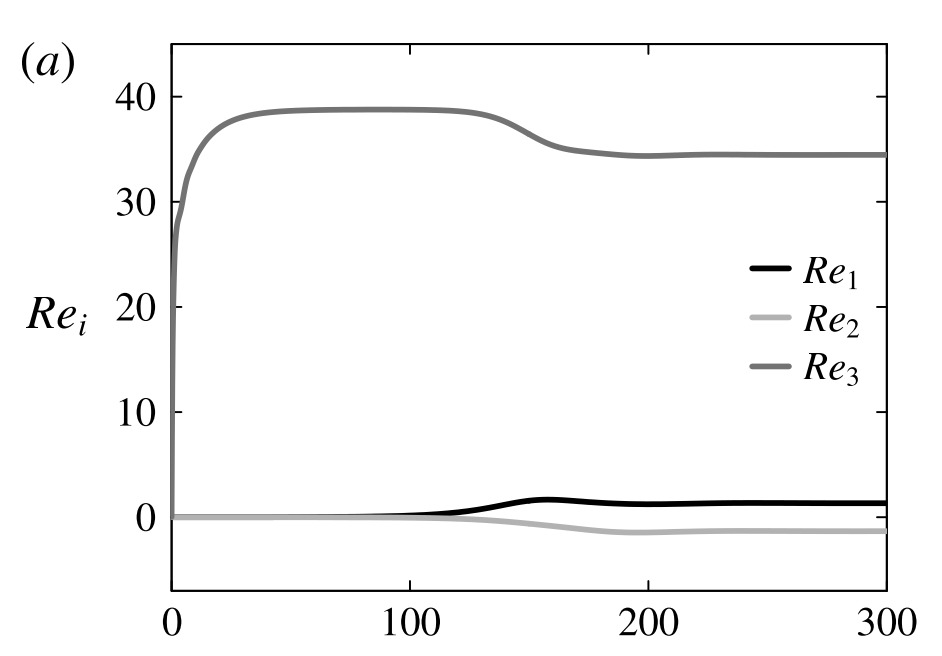}} \\
\caption{Time evolution of two components of the bubble Reynolds number for two different cases of steady oblique rise.
Present results on the left panel, DNS by~~\citep{loisy2017buoyancy} on the right panel.}
\label{Fig:3Doblique-a}
\end{figure}

The last set of moderate Re test cases is the oblique rise of periodic arrays of bubbles performed by ~\cite{loisy2017buoyancy}. 
The numerical experiment consists in initializing a perfectly regular lattice of bubbles and following its evolution. 
%The lattice should be large enough (in principle infinite) to minimise boundary effects. 
%From a numerical point of view, 
As before, using periodic boundary conditions it is possible to simulate an infinite lattice with a single bubble. 
The volume fraction can be varied by changing the size of the bubble relative to the cell size.
~\citet{loisy2017buoyancy} pointed out that for certain values of the non-dimensional parameters, bubbles can have an oblique trajectory (not aligned with gravity) at certain volume fractions, although a single bubble in the same parameter regime would follow a straight vertical path.
Analytical considerations support the possibility of a non-trivial path indicating a possible transition for $Ar\approx 20$. 
In particular three different oblique regimes have been found: (a) a steady oblique rise, (b) an oscillatory oblique rise, with a bubble oscillating around a straight oblique path, and (c) a chaotic oblique rise.
Such a behaviour had been previously noticed numerically ~\citep{sankaranarayanan2002analysis}, but using a diffuse interface method and a small density ratio. 
In the present work we have simulated the configurations corresponding to the three regimes in ~\cite{loisy2017buoyancy} with a slightly increased resolution, to test if our numerical approach confirms these previous results.
The density ratio and the viscosity ratio between the two phases are the same for all the cases, $\rho_b/ \rho_l = 0.005$, $\mu_b / \mu_l = 0.01$. The number of points is varied with the domain size in order to always get the same bubble resolution $d_b / \Delta = 40$.
Global agreement between present simulations and those by ~\cite{loisy2017buoyancy} is excellent.
In Fig. \ref{Fig:3Doblique-a} we show as an example 
the comparison for r\'egime (a). 
The details of these simulations together with other results are given in Appendix \ref{app:array}.

Before analysing complex flows at high Reynolds numbers, we have also carried out a specific quantitative analysis on the effect of two crucial numerical issues: (i) resolution, (ii) the density ratio.
It is worth emphasising that there is a strong link between physical properties and numerical parameters and that this cannot be overlooked.
While the simulation of a single bubble remains feasible 
even with a very fine grid thanks to the adaptive mesh, it would not be possible to tackle a problem with many bubbles with the same grid. 
Moreover without the adaptive mesh even the single bubble case appears desperate at large Reynolds numbers.
In contrast, using a coarse grid may make the computation easy but the results might be largely unreliable.
We summarise here our findings, details are given in the Appendix.
In order to simulate bubble flows quantitatively and in detail, it turns out to be key: 1) to have a large number of points per diameter (we have found that convergence is obtained with about $N_{\rm points}\approx Ar/2$);
2) using an adaptive mesh, it is sufficient for fine cells to be concentrated inside the bubble and in the wake; 3) a large density ratio is mandatory to avoid spurious effects which are similar to those found with a coarse grid. Namely, $\rho_l/\rho_b > 100$.
Our results confirm the previous analysis by ~\cite{cano2016global}.
%Our results therefore point out that to reproduce multiphase flows, a realistic density ratio is mandatory, while it possibly may be slightly less than the actual one.

%\clearpage

\section{Pseudo-turbulence in two dimensions}
\label{sec:2D}
\begin{figure}
\center
{\includegraphics[width=.25\textwidth]{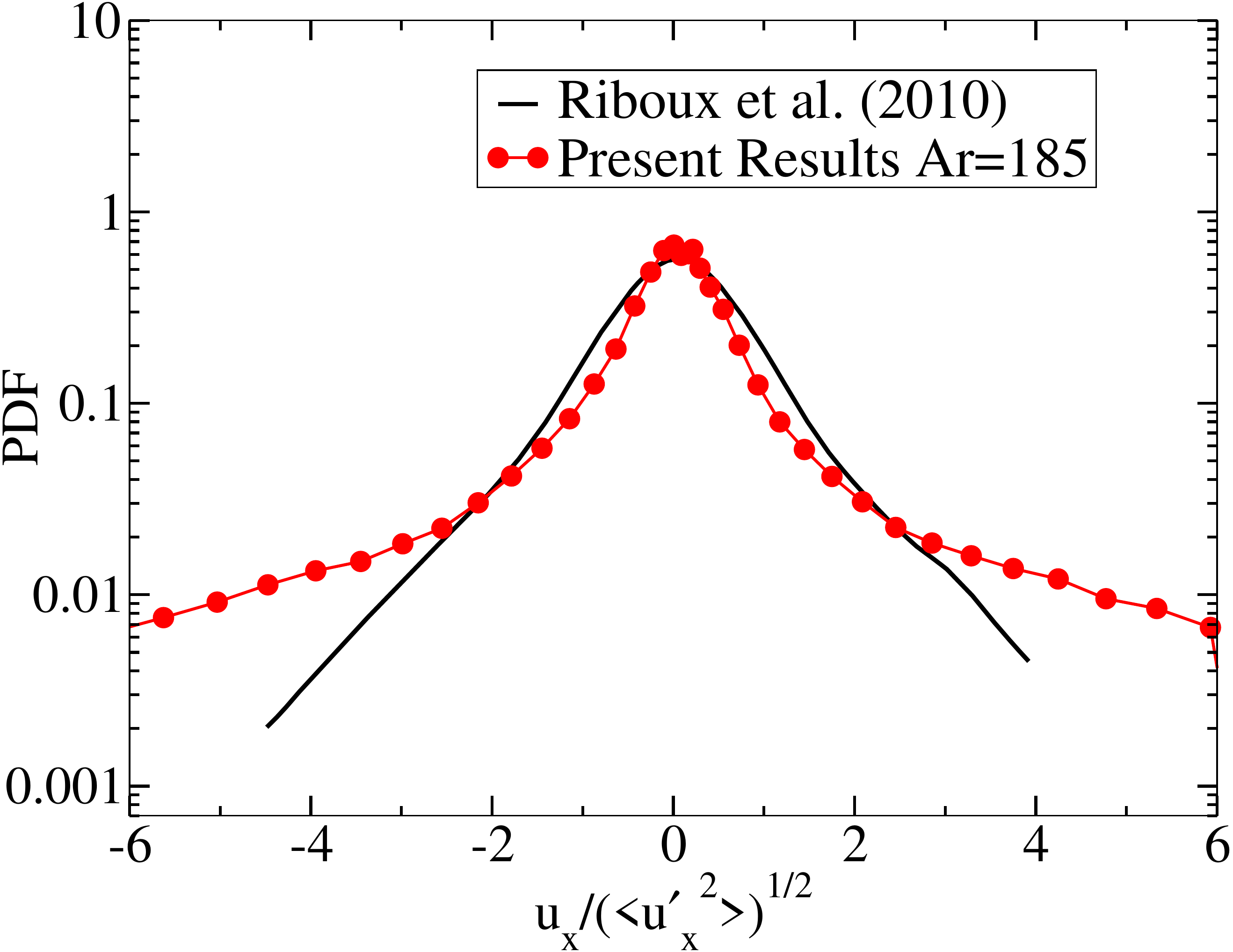}} 
\hspace{0.5cm}
{\includegraphics[width=.25\textwidth]{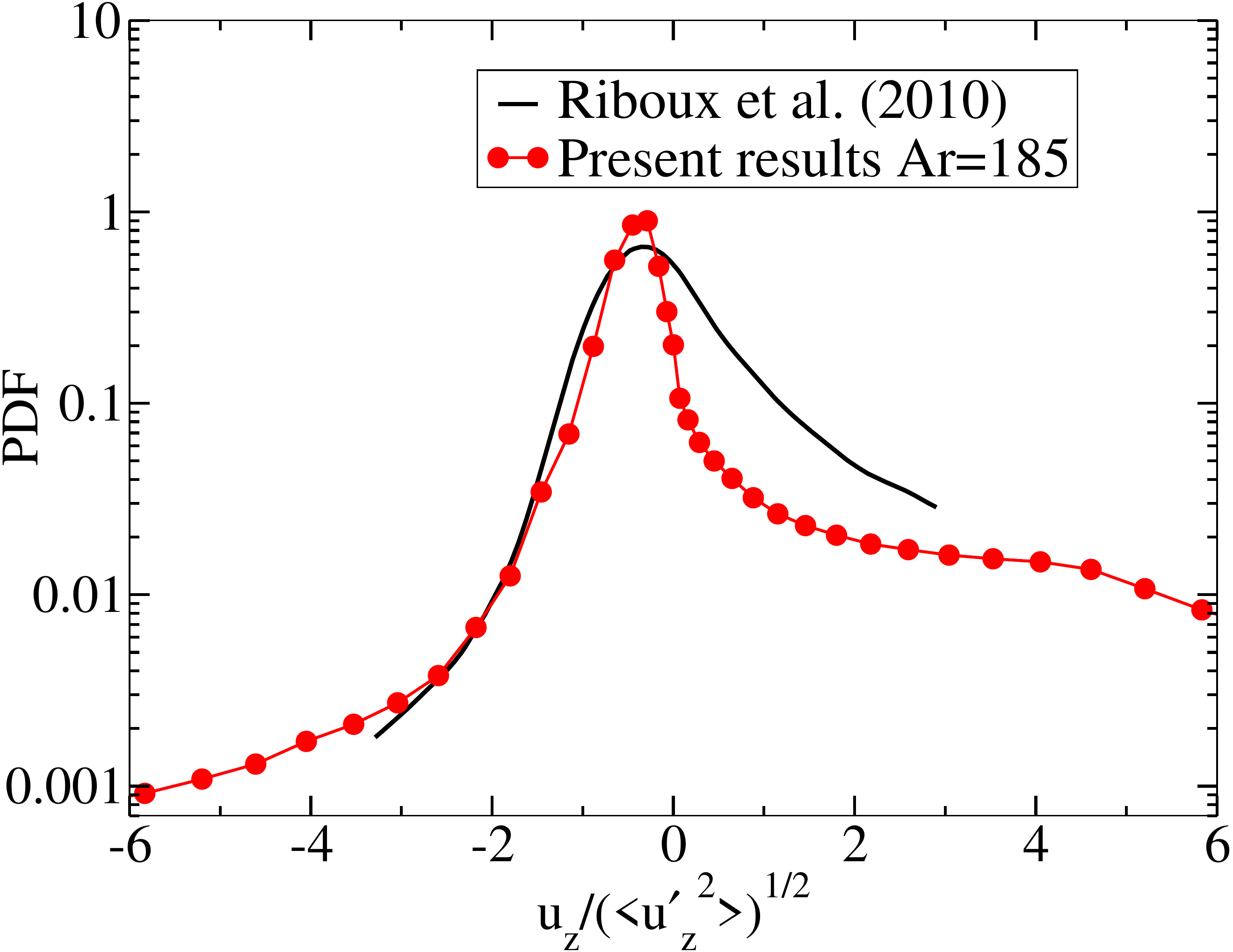}}
\caption{Bubble positions within the domain at $t=0$ for simulation (a).  Left: a sketch. Right the numerical vorticity field at t=0.4.}
\label{Fig:init}
\end{figure} 
In this section we discuss the results of a 2-dimensional bubble column configuration. We consider a square domain with the vertical direction $z$ aligned with gravity, acting downward.  The tank, of size $50 d_b\times 50 d_b$,  is filled with a liquid and $32$ initially spherical bubbles are placed at the bottom, in a region confined between $z = 0$ and $z = 8 d_b$, and are homogeneously distributed in the lateral direction $y$, while avoiding any initial bubble overlap, and with a minimum distance between them of one diameter. This results in a local volume fraction in the region $0 \le z \le 8$ of $\alpha \simeq 5 \%$. The domain is closed at the bottom by a wall (no-slip boundary condition), and an outflow boundary condition is used at the top, while on the lateral sides the domain is periodic. At $t=0$ both the liquid and the bubbles are at rest. A picture of the initial condition is shown in figure \ref{Fig:init}.
\begin{table}
\begin{center}
\begin{tabular}{ c  c  c  c c}
\hline
%\toprule
Case & $Ar$ & $Bo$ & $N$ & $d_b / \Delta$ \\
\hline
%\midrule
$a$ & $100$ & $0.12$ & $4096$ & $82$\\
$b$ & $140$ & $0.20$ & $8192$ & $164$\\
$c$ & $313$ & $0.56$ & $16384$ & $328$\\
\hline
%\bottomrule
\end{tabular}
\caption{Non-dimensional parameters for the 2-dimensional bubble column. $N$ represents the grid resolution.}
\label{tab:bub_column}
\end{center}
\end{table}
The flow parameters are given by the non-dimensional Archimedes and Bond numbers, defined in the previous sections. 
The viscosity and density ratios are constant in all the simulations and their values are $ \mu_l / \mu_b = 100$ and $\rho_l / \rho_b = 1000$. 
%{\color{red} SP: I don't like it when non-dimensional parameters are mixed with dimensional parameters. We have kept constant the values of the surface tension and gravity, and we have varied the bubble diameter to investigate different regimes.
% Specifically, the physical parameters are very close to those of air and water, having been fixed $\sigma = 0.05 \, N/m$ and $\mu_l = 7 \times 10^{-4} \, Pa$. 
% Varying the bubble diameter from $d = 0.8 \, mm$ to $d=1.7 \, mm$,}
 Three different simulations with bubbles of different diameters have been carried out, and the corresponding parameters are reported in table \ref{tab:bub_column}.
In particular, the $Ar$ number is within the range $Ar \simeq 100 - 300$, which corresponds to typical three-dimensional experiments ~\citep{riboux2010experimental}.
%Given that we consider a true 2D configuration,
%the regimes are possibly quite different from the 3D case
%with respect to the path instability of the single bubble ~\citep{tchoufag2014linear,cano2016global}.
%Inspecting the trajectories, it appears that the cases (a) and (b) are within the stable regime, 
%in which the trajectory of the undisturbed bubble is rectilinear.
%Case (c) is more complex and the trajectory is chaotic.
%{\color{red}SP: this is 2D, no? so the regimes are possibly quite different, and never spiral anyway}.
From a numerical point of view, in all the three cases we have used regularly spaced grids with different resolutions depending on the increasing bubble Reynolds number. 
A regular grid has been preferred to the adaptive one since in two dimensions the gain with adaptivity is not as significant as in 3D, and because a regular grid spacing facilitates the computation of statistical measures (spectra in particular).  
In any case, the resolution requirements to get physically-sound results have always been fulfilled, as highlighted in table \ref{tab:bub_column}.

%\subsection{Energy spectra}
%\begin{figure}
%{\includegraphics[width=.48\textwidth]{fig10a.eps}} 
%{\includegraphics[width=.48\textwidth]{fig10b.eps}} \\
%\caption{Velocities of the bubbles with respect to time. On left panel case (a), on right panel case (b).}
%\label{Fig:velocity_bub}
%\end{figure} 
%In figure \ref{Fig:velocity_bub}, we display the velocity of the bubbles as a function of time for cases (a) and (b).
%It can be observed that after a transient, of the order of $t_{\rm trans}\sim 10$, the bubble swarm spreads and is stretched in the vertical direction.
%Bubbles are in general disturbed in their motion and some bubbles are affected by high levels of fluctuations with respect to the mean velocity of the front. Notably, a few bubbles even remain trapped in the wake region with a significantly lower vertical velocity. This is due mainly to the fact that 2D interactions are stronger and have a longer range than in 3D. This results in a more disturbed path of the bubbles in the wakes which can deviate significantly from vertical motion with non-trivial downward motion at some instants.                                                                                                                                                                                                                                                      
We first consider the energy spectra. 
Since the problem is non-homogeneous in the vertical direction and non-stationary, particular care must be taken in the definition of the interrogation window. In the experiment of ~~\cite{riboux2010experimental}, spectra are evaluated after the passage of the bubble swarm in a centred square box.
The spectra $S_{ii} = \langle | \hat{u}_i |^2 \rangle$, where $\hat{u}_i$ is the Fourier transform of the velocity fluctuation in the $i$ direction, are evaluated separately for the vertical and the horizontal components. The transform is performed in the $y$ direction, which can be considered as statistically homogeneous, for both components of the velocity.
To improve statistics we have also averaged in the $z$ direction over windows of length $5d_b$, which have been found to be statistically homogeneous to a good degree of approximation.
%Spectra obtained for case (a) are displayed in figure \ref{Fig:spettri2}.
%In all figures the energy is made non-dimensional with the corresponding standard deviation and on the $x$ axis we show the wave number $k = 2\pi n/50$ with $n=1,..N$. The length is normalised with the bubble diameter, and the value corresponding to the bubble diameter is indicated by the vertical line.

\begin{figure}
%{\includegraphics[scale=0.5]{fig11a.eps}} 
{\includegraphics[width=0.5 \textwidth]{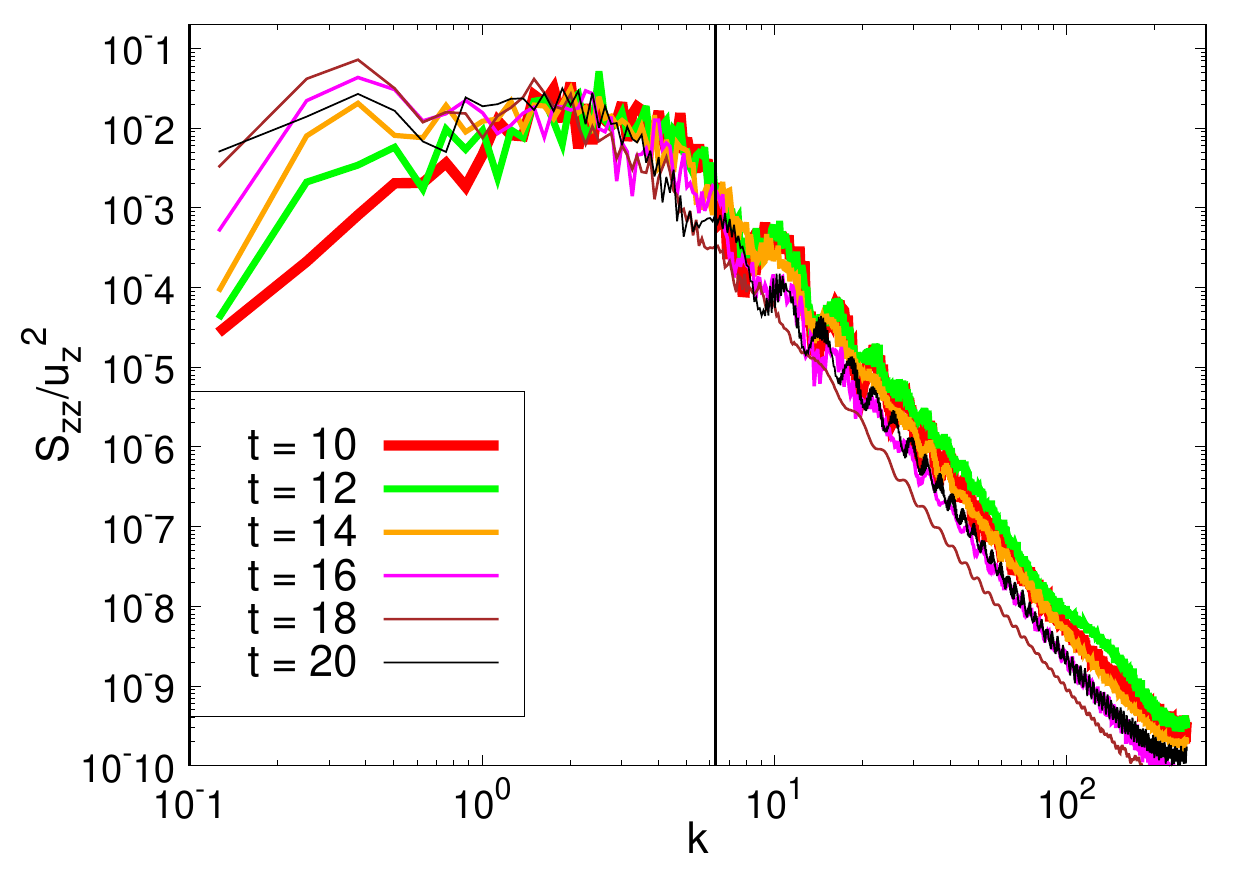}} 
%{\includegraphics[scale=0.5]{fig11c.eps}} 
{\includegraphics[width=0.5 \textwidth]{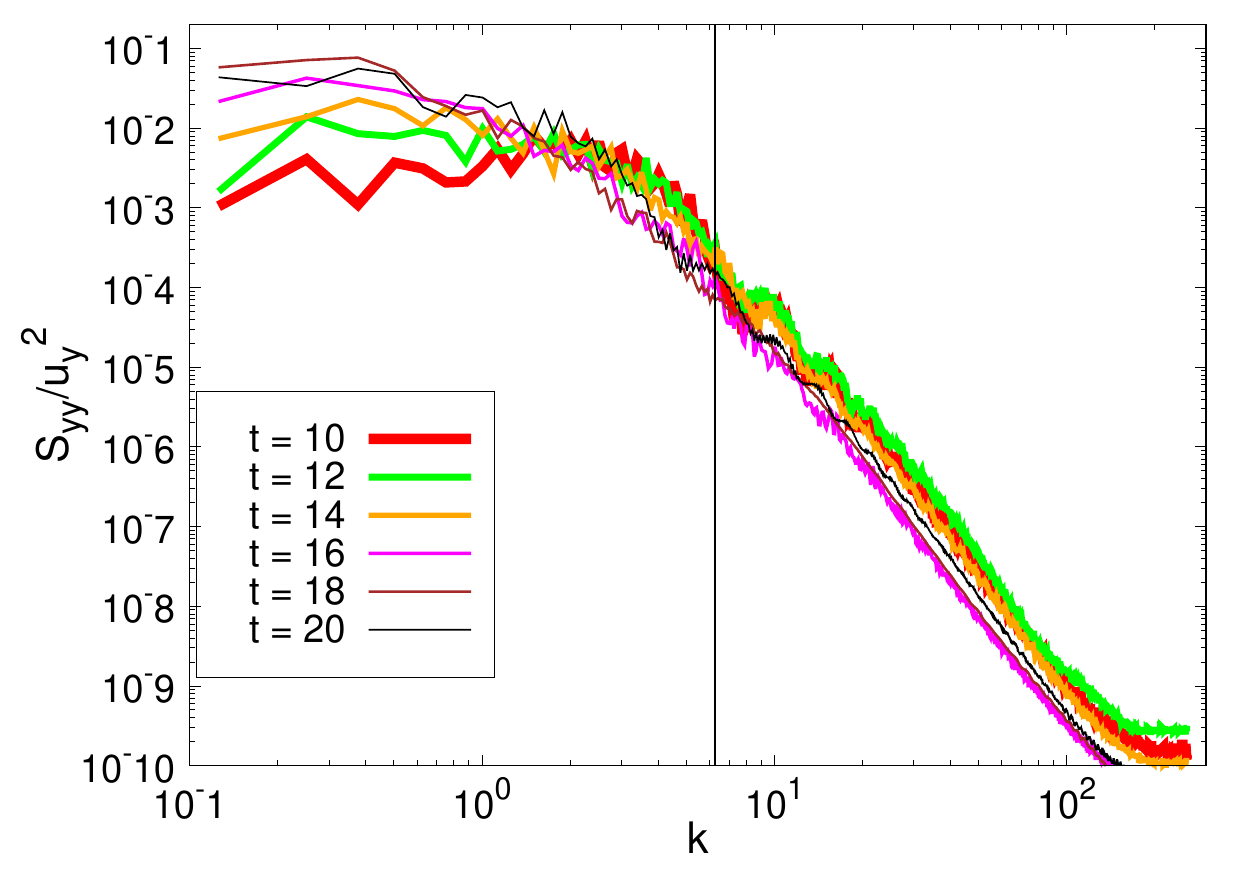}} 
\caption{Spectra of the vertical (top line) and horizontal component (bottom line) of the velocity for bubbles with $Ar = 100$ and $Bo = 0.1$ evaluated at different times. 
The spatial interrogation window is between  $20$ and $25$. 
%The dashed line represents the $-3$ slope. 
The energy spectrum is made non-dimensional with the corresponding standard deviation.
The vertical line corresponds to the bubble diameter.}
\label{Fig:spettri2}
\end{figure} 
In figure \ref{Fig:spettri2}, we show the spectra of both the vertical and horizontal velocity fluctuations evaluated at different times, made non-dimensional with $\sqrt{d_b/g}$, for the case at lower $Ar$.
It can be seen that the spectra are independent from time, over the whole time-window considered, that is from the instant when all the bubbles have entered the spatial $z=15 d_b$ plane.
%This time spans from when all the bubbles has entered the spatial interrogation window and the time at which they have all left the region.
In particular, the spectral slope appears rather constant. 
Furthermore, no appreciable difference is found between the horizontal and the vertical spectrum, showing that both components dynamically distribute the energy in a similar manner.
Then, we can write that the 1D spectrum is $E(k)=S_{ii}$ without compromise.
The results obtained at the other $Ar$ numbers show the same trend, and are not displayed for the sake of simplicity.

\begin{figure}
\center
{\includegraphics[scale=0.25]{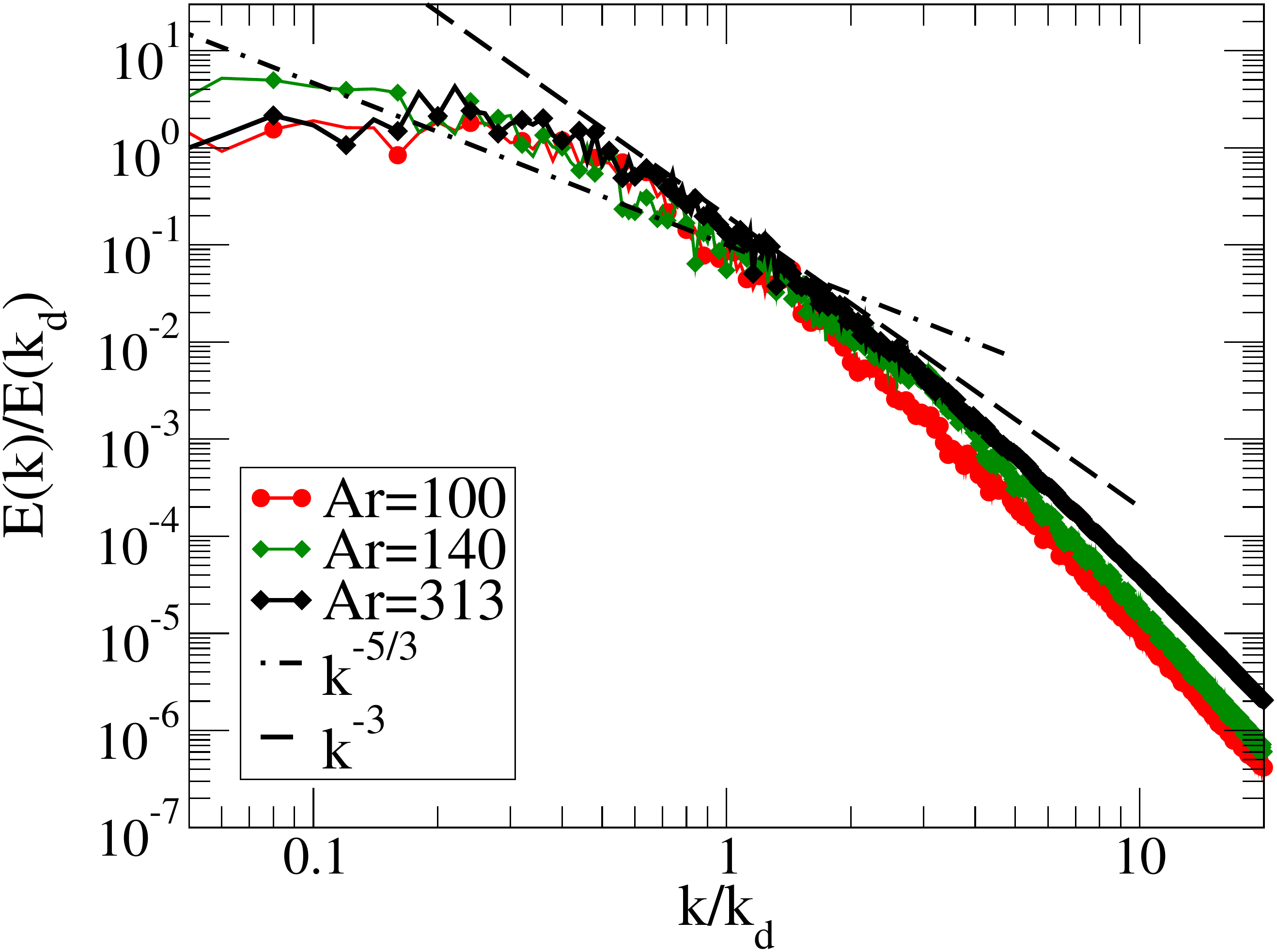}} 
%{\includegraphics[scale=0.5]{fig5a2.eps}} 
{\includegraphics[scale=0.25]{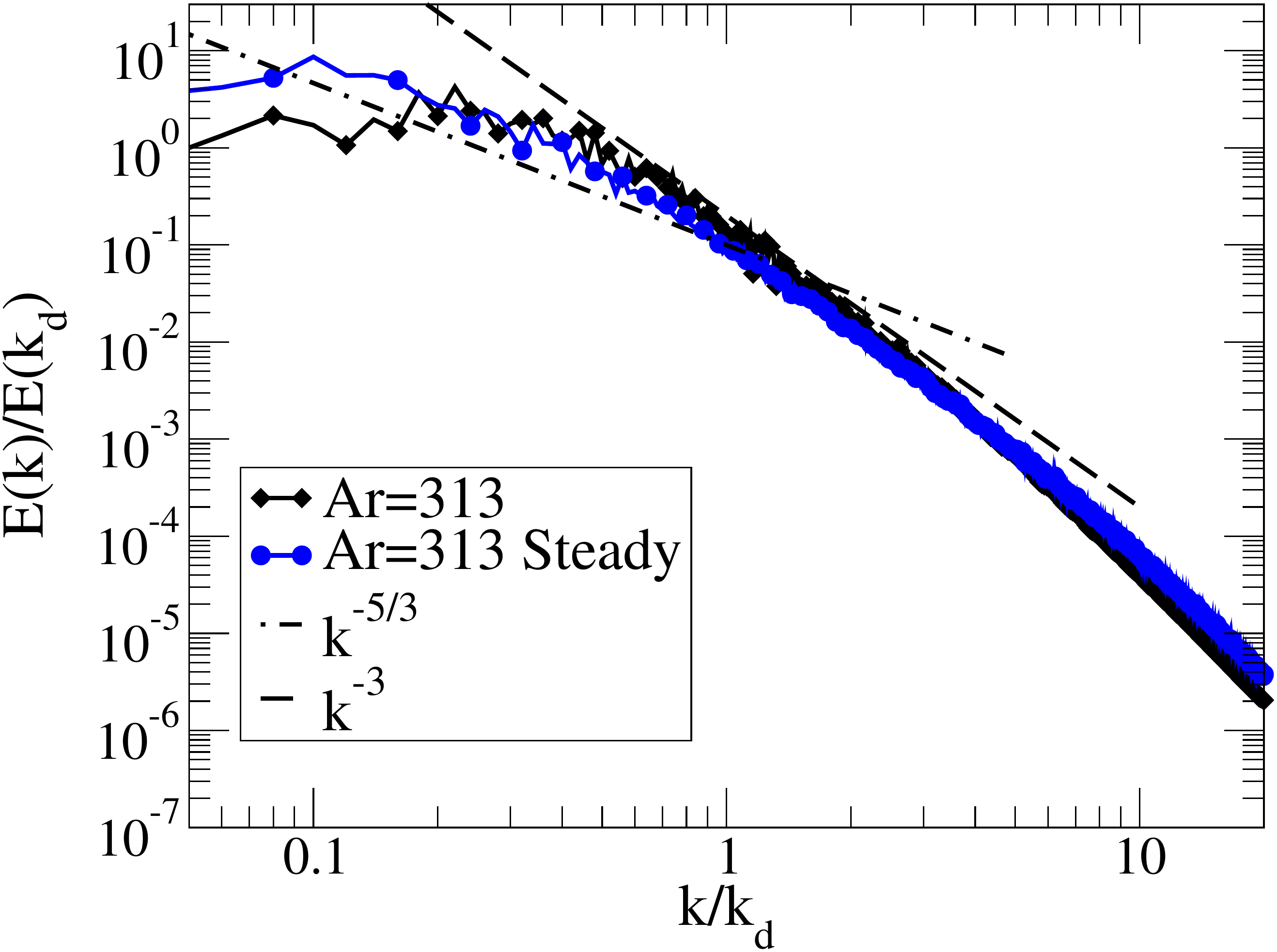}} 
%{\includegraphics[scale=0.5]{fig5b2.eps}} 
\caption{(a) Spectra of the vertical component of the velocity of bubbles for different $Ar$ evaluated at the time $t=15$.
The vertical line corresponds to the bubble diameter. 
The dot-dashed line indicates the $k^{-5/3}$ slope, and the dashed line the $k^{-3}$ slope.
(b) Energy spectrum of vertical fluctuations against $k$ for the case $Ar=313$ for both the unsteady and the steady configurations.
For the steady case, the spectrum is obtained by averaging over time between $t=13$ and $t=23$. Lines are the same as (a).}
\label{Fig:spettri3}
\end{figure} 
%\begin{figure}
%\center
%{\includegraphics[height=5cm]{fig21a.eps}} 
%%{\includegraphics[height=4.5cm]{fig21b.eps}} 
%\caption{ Energy spectrum of vertical fluctuations against $k$. The spectrum is obtained by averaging over time between $t=13$ and $t=23$. The vertical line corresponds to bubble diameter. 
%The dot-dashed line indicates the $k^{-5/3}$ slope, and the dashed one the $k^{-3}$.
%%(b) Mean energy flux, Eq. (\ref{eq:flux}), with different filter lengths. The time-average is taken in the range $t=13-23$ as in (a).
%}
%\label{Fig:periodic-spettri}
%\end{figure} 
We then compare the spectra at different $Ar$ numbers, at the same time $t=15$, as shown in figure \ref{Fig:spettri3}.
In all cases spectra are compatible with a scaling $E(k)\sim k^{-3}$ in a range around the diameter scale. 
At small scales, a steeper scaling $E(k)\sim k^{-4}$ is found also in all cases, which can be related to a range where viscous effects are important \citep{Mon_75}.
However for the case (a) this dissipative range appears to dominate over almost the whole range of scales smaller than the diameter.
In case (b), the spectrum displays a $-3$ slope over roughly a decade, while for the highest $Ar$ number the range appears even larger.
Moreover, we observe for the (b) and (c) cases that around the bubble diameter there is a crossover and the spectrum is flatter with a slope close to $-5/3$.
%Yet the separation of scales is too small to draw a definite conclusion about this range.
To check the statistical robustness of our analysis, we have repeated the simulation of the case at $Ar=313$ with periodic conditions in both directions.
In this case, the flow is statistically homogeneous in all directions, and after a transient a steady state is attained.
Therefore both spatial and time averages are taken.
%In figure \ref{Fig:periodic-spettri}a, we show the spectrum of vertical fluctuations and we can see that 
The periodic simulation 
confirms the results obtained in the unsteady case. In particular, a $k^{-5/3}$ scaling is obtained at  scales larger than the bubble diameter. The $k^{-3}$ scaling appears to be present at scales smaller than the bubble diameter and then a steeper slope typical of a viscous range is found.

\begin{figure}
\center
{\includegraphics[scale=0.2]{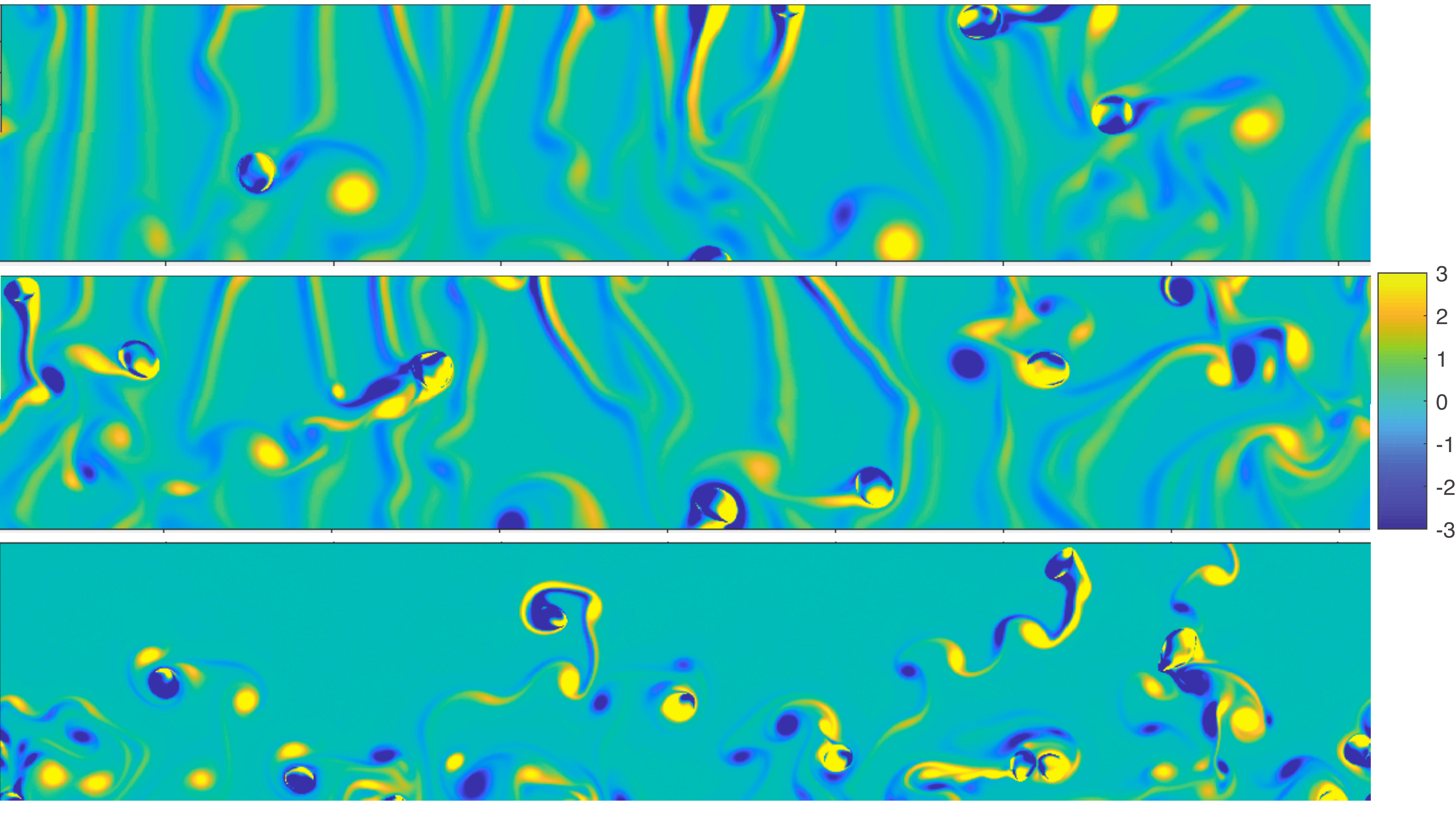}}
%\begin{tabular}{cc}
% {\includegraphics[scale=0.5]{Ar100b}} & \\%\vspace{-0.95cm}  
% {\includegraphics[scale=0.5]{Ar100a}}  &\\ 
% {\includegraphics[scale=0.5]{Ar140b}} &\\ %\vspace{-1.cm}  
% {\includegraphics[scale=0.5]{Ar140a}} & {\includegraphics[scale=0.4]{colorbar.eps}}\\
% {\includegraphics[scale=0.5]{Ar313b}} & \\
% {\includegraphics[scale=0.5]{Ar313a}} & 
% \\
 %{\includegraphics[scale=0.8]{fig16b.eps}} &  %{\includegraphics[scale=0.35]{colorbar.eps}}
%(c) &  {\includegraphics[scale=0.58]{fig12c.eps}} 
%{\includegraphics[scale=0.58]{colorbar.eps}}\\
%\end{tabular}
\caption{The Vorticity field displayed at $t = 16$ and in the domain between $15$ and $25$ bubble diameters in the vertical direction, for the different $Ar$ cases:
(a) $Ar = 100$ and $Bo = 0.1$;
(b) $Ar = 140$ and $Bo = 0.2$;
(c) $Ar = 313$ and $Bo = 0.33$.
The colour bar is the same for the three cases and is displayed laterally.}
\label{Fig:vorticity1}
\end{figure} 
In Figure \ref{Fig:vorticity1} we show the vorticity field in the two portions of domain that have been used for the evaluation of the spectra at a fixed time $t = 15$. 
The visualisation allows to link the statistical spectral properties to the actual dynamics of the flow.
The bubbles are a source of vorticity which then creates the trailing wakes.
Some persistent coherent structures are clearly visible.
We observe that at  $Ar=100$, the interaction between the wakes exists but is small, notably in the upper part of the window. 
%This suggests that spectra are dominated by the coherent structures generated around the bubbles at the diameter scale for this case.
The plot for $Ar=140$ clearly suggests a stronger interaction between bubble wakes. 
We can observe that the vorticity field is diffused through non-linear interactions. Because of them, some bubbles are even led to go backward against gravity.
The case at $Ar=313$ is similar to the $Ar=140$, but the strong interaction between wakes and the presence of dynamics at smaller scales are even more visible, with thin unstable vorticity filaments released behind the bubbles.
The nonlinear wake interactions are clearly dominant here and bubbles follow quite intricate paths.
% and 
%that the spectra scaling-law that has been found has little relation to the structures of the wakes, since in the upper window no significant interaction between the wakes is present, even though the resulting spectra are very close to those obtained in the lower window.
Although a $k^{-3}$ scaling has been found in all cases, the present results show that in case (a)  the spectrum is basically related to the coherent structures of the wakes. 
In contrast for the other two cases,  because of the higher Reynolds number,  turbulent mechanisms play an important role. 
Notably bubble dynamics lead to an injection of energy and vortiticy at the scale of the bubble diameter and
 energy is transferred towards different scales through non-linear mechanisms.
 In both cases at $Ar=140$ and $Ar=313$ these interactions are significant enough that energy is transferred to larger scales by an inverse cascade, which is revealed by the corresponding $-5/3$ scaling of the spectrum.
The steeper range $E(k)\sim k^{-4}$ found in all cases at small wavelengths, can be linked to the viscous damping regime. The typical dynamical features of this regime can be seen in the bottom portion of the domain of case (a).
%In particular, the bubbles inject much energy and the behaviour of the spectrum at the largest scales for this case should be related to a 2D inverse cascade.
%The importance of interactions can be inferred also by the fact that, at variance with previous cases, no exponential viscous decay begins in the time window considered, and bubbles need more time to form the spectrum in the higher window. 
%In fact, because of the strong interactions, bubbles are slowed down by their turbulent vortical dynamics. 
%The intermediate scaling, which possibly reveals a $-3$ scaling may be associated to the presence of vorticity at small scales generated in the wakes.
%{That might in turn trigger a direct cascade of enstrophy.}  

%\begin{figure}
%\caption{PDFs of the velocity fluctuations in the vertical $zz$ (left panel) and lateral $yy$ (right panel) directions. Bubbles with $Ar = 313$ and $Bo = 0.33$. The window is between $z=15d_b$ and $z=20d_b$.}
%\label{Fig:pdf2}
%\end{figure} 
\begin{figure}
{\includegraphics[scale=0.25]{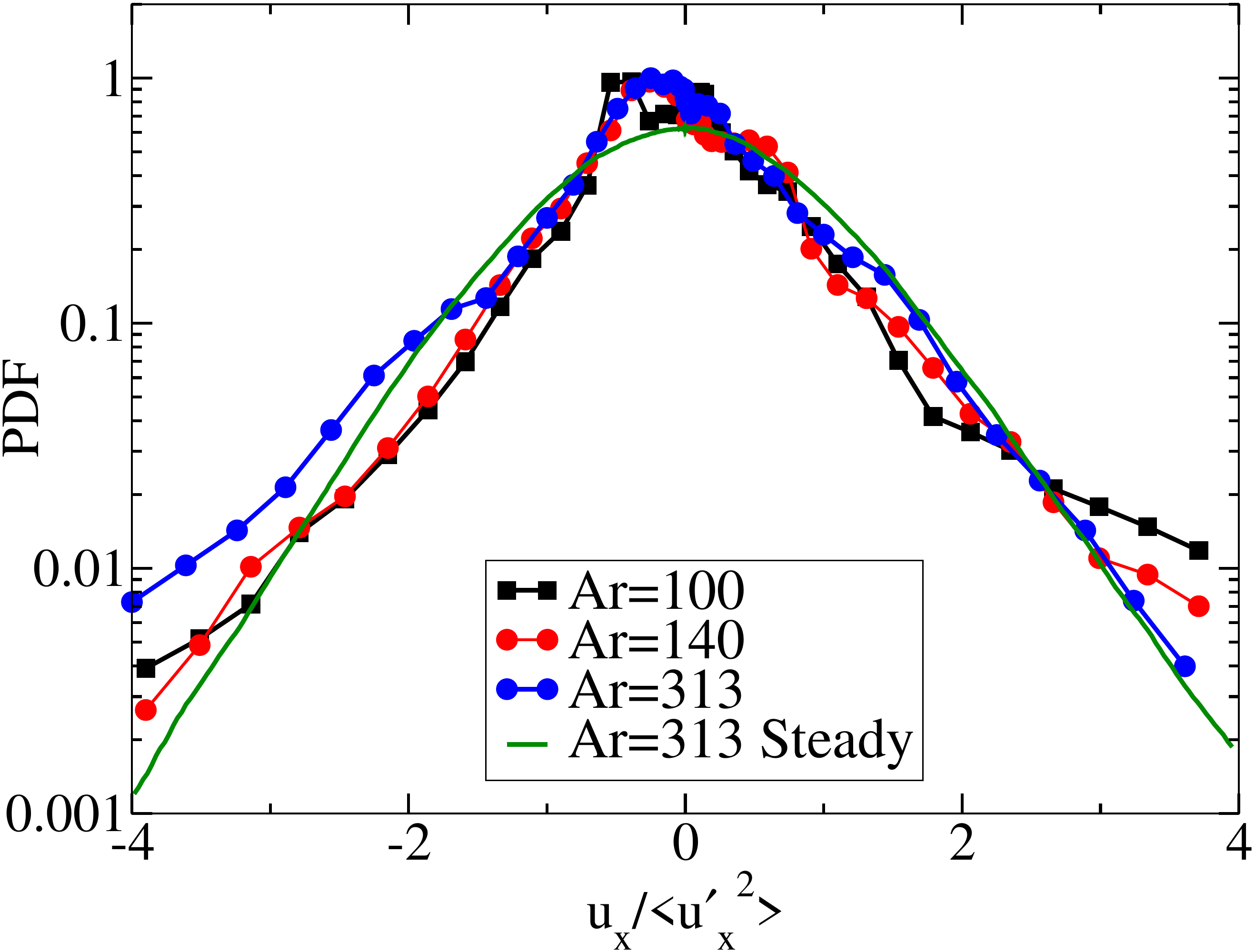}} 
{\includegraphics[scale=0.25]{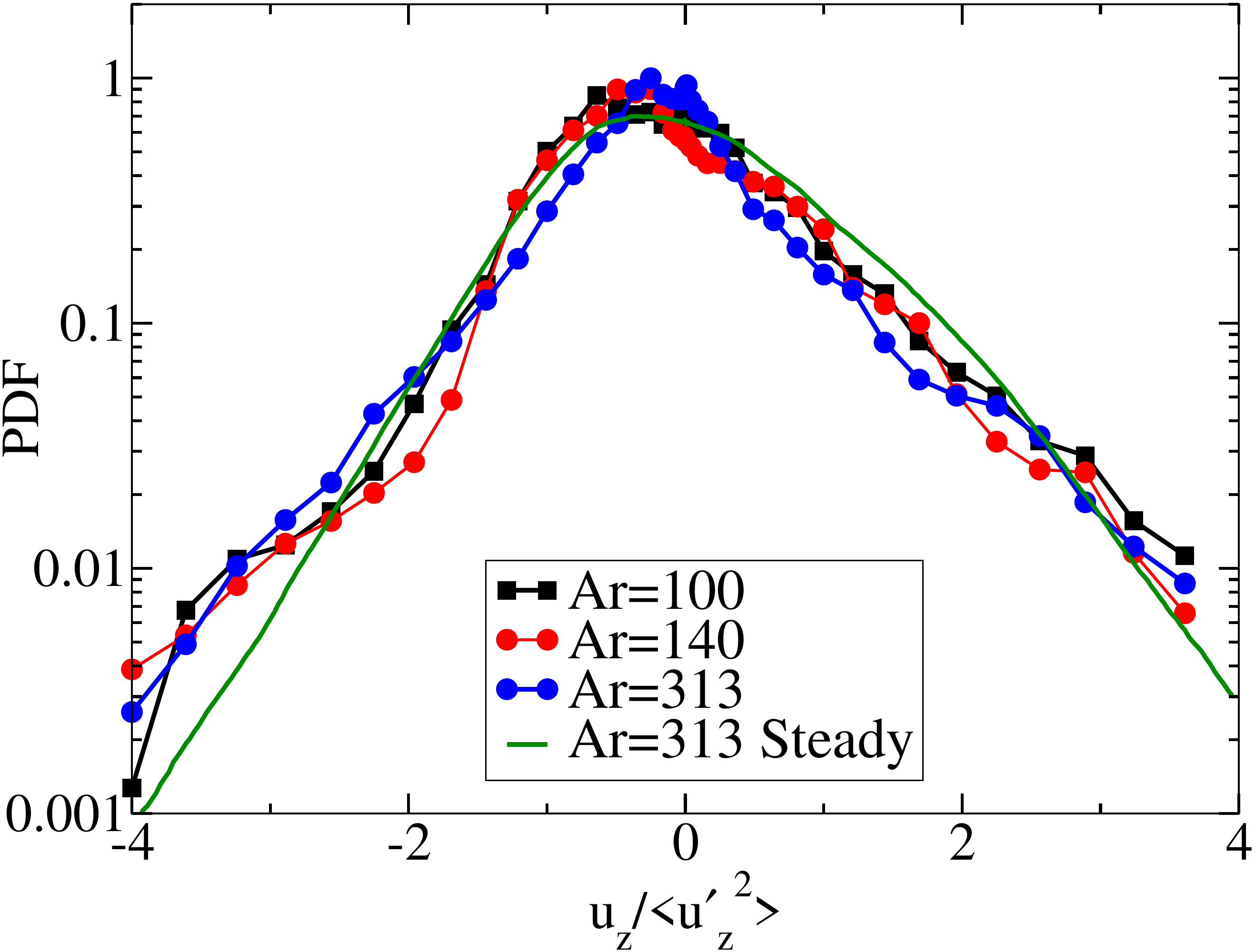}}\\
\caption{PDFs of the velocity fluctuations in the vertical $zz$ (left panel) and lateral $yy$ (right panel) directions for different $Ar$. The steady simulation at $Ar=313$ is plotted for comparison.
As before, in the steady case, average in time has been taken in the window $t=13-23$. }
\label{Fig:periodic-pdf}
\end{figure} 
In figure \ref{Fig:periodic-pdf}, the PDFs of the velocity fluctuations for the different cases are shown together with those obtained in the steady case.
From a physical point of view, PDFs are clearly not Gaussian with exponential tails, and while the horizontal one is symmetric, the vertical one is skewed, showing anisotropy of fluctuations and the particular status of the vertical direction.
From a statistical point of view, the PDFs show unambiguously that results obtained in the unsteady regime are statistically robust, provided the analysis on the liquid is performed when bubbles have passed through it.
In our case, this happens at about $t=13$ for all $Ar$ numbers. 
After that time, results are basically frozen for some characteristic times.
Of course, smoother profiles are obtained in the steady case because of the time-averaging.

In conclusion for this section, we have exhaustively investigated the 2D case focusing in particular on the proper numerical computation of statistics. We can draw the following conclusions:
\begin{itemize}
\item In our numerical simulations we can disentangle between the gas and liquid phase, and have checked that in our configuration there is no appreciable difference if statistics are computed when bubbles are present or not, consistently  with what was already pointed out by~\cite{dodd2016interaction} for droplets.
\item The flow is always statistically homogeneous in the horizontal direction, and can be considered homogeneous in the vertical direction also in the unsteady case, on a scale of a few diameters. Given the high time-resolution, the flow can be considered steady (frozen) over some timesteps even in the unsteady case, in the central region of the column.
\item Both in unsteady and steady simulations the transient regime has to be left out from the statistics. This corresponds to about 10 diameters from the bottom in the unsteady regimes investigated here. When the transient is left out the statistical analysis is robust and results both for one-point (PDFs) and two-point (spectra) observables are in agreement with the same statistics computed in the steady state.
\item On the basis of the spectral analysis, we can reach the conclusion that, at least with concentration around $5\%$, the $Ar$ number must be higher than $100$ to trigger significant interactions and turbulent-like cascades, even in the 2D case where interactions are stronger than in the 3D case. This conclusion has been confirmed by the analysis of the other statistical features, namely the PDF of the velocity fluctuations and the flux of turbulent energy.
\item From a physical point of view, some results are interesting, although the 2D configuration is not realistic: (i) it has been highlighted that in 2D an inverse cascade is triggered from the injection of energy at the scale of one diameter; (ii) The $k^{-3}$ spectrum scaling has been found in all simulations, but our simulations show clearly that the same spectrum can be related to different mechanisms, namely the fluctuations around wakes \cite{risso2011theoretical} in the moderate $Ar$ flow, or to a nonlinear cascade of turbulent energy.
\end{itemize}

This results have been used to build up the 3D simulation described in the following section.
Some more details can be found in the Appendix.

\section{Three dimensional bubble column}
\label{sec:3D}
\begin{figure}
\center
{\includegraphics[width=.7\textwidth]{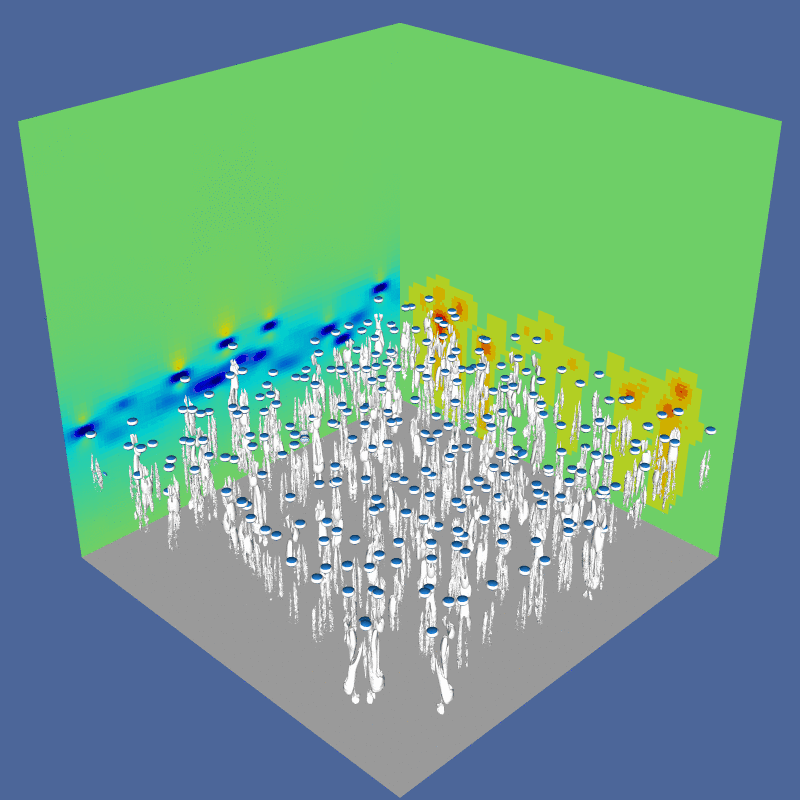}} 
%{\includegraphics[scale=0.58]{./spettri/N140/pdf_y_t=10-20_N140.eps}} \\
\caption{Snapshot of the 3D simulation at $t=6$ after the release. The VOF field is showed with blue iso-surfaces and the $\lambda_2$ vorticity field is shown with grey contours. The right wall displays the level of mesh adaptation, while the left panel displays the vertical component of the velocity field.}
\label{Fig:3D}
\end{figure}
The 3D bubble column is a direct extension of the previous 2D numerical experiments:  
the cubic tank, of size {$50 d_b \times 50 d_b \times 50 d_b$},  is filled with a liquid and $256$ initially spherical bubbles are placed at the bottom, in a region confined between $z = 0$ and {$z = 3 d_b$}. The bubbles are homogeneously distributed in the lateral directions $x,~y$, while avoiding any initial bubble overlap, and with a minimum separating distance of $1$ diameter. 
This results in a local volume fraction in the region $0 \le z \le 3$ of {$\alpha \simeq 2 \%\div3\%$}. 
The domain is closed at the bottom by a wall (no-slip boundary condition), and an outflow boundary condition is used at the top, while on the lateral sides the domain is periodic. 
At $t=0$ both the liquid and the bubbles are at rest. 
The dimensionless characteristic numbers of our numerical experiment are the following:
$Ar=185~;~Bo=0.28~; {\rho_l/\rho_b=800}~;~\mu_l/\mu_b=100$.
The configuration is in many respects equivalent to that investigated experimentally in three dimensions by ~\cite{riboux2010experimental}.

From the numerical point of view, an adaptive mesh has been used with a maximum possible refinement of $N=4096$ cells in each direction, meaning a maximum resolution in terms of the bubble diameter of $d_b / \Delta = 82$. 
The grid is refined or coarsened relying on the errors on the volume fraction and on the velocity components, using as absolute thresholds for the refinement the values $e_f = 0.01$ and $e_v = 0.003$, based on a previous analysis that is detailed in Appendix \ref{app:tech}.
The total number of computational cells grows in time because of the elongation of the wakes, starting from $N_{tot} \simeq 10^7$, and attaining  $N_{tot} \simeq 9 \cdot 10^8$ at $t= 12$. 
%Consequently, also the number of CPUs is increased in time to maintain a well balanced parallelisation, attaining a final number of $N_{cpu} = 1536$. 
Note that using a non-adaptive mesh would require $4096^3\approx 69\times 10^9$ grid points, which is  beyond present computational capabilities.
The present numerical experiment is therefore basically the best that can be done in simulating bubbly flows today.

At a qualitative level, Figure \ref{Fig:3D}
shows the instantaneous motion of the bubbles at time $t=6$, made non-dimensional with the bubble buoyancy time $\sqrt{d_b/g}$. The vorticity generated by the bubbles is included in elongated wakes. 

\begin{figure}
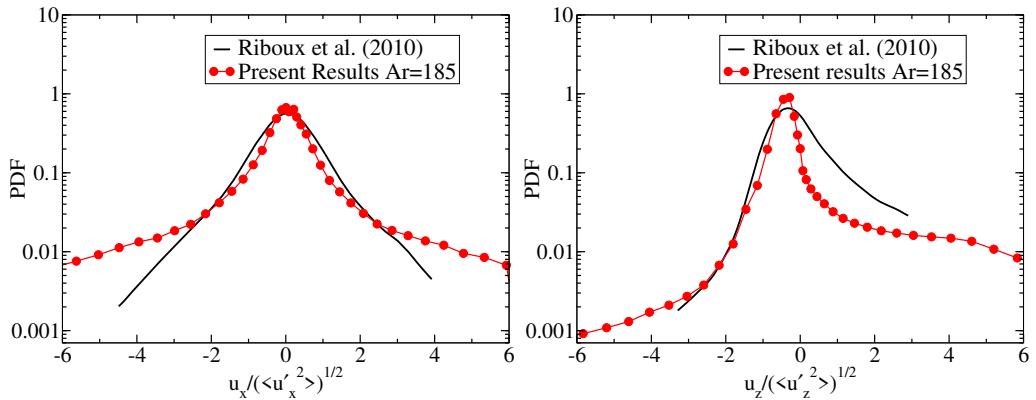

{\includegraphics[scale=0.25]{fig9a.eps}} 
%{\includegraphics[scale=0.5]{fig28b.eps}} \\
{\includegraphics[scale=0.25]{fig9b.eps}} 
%{\includegraphics[scale=0.5]{fig28d.eps}} \\
\caption{PDFs of the velocity fluctuations in the lateral $x$ direction ($u$) (a), and in the vertical $z$ (b), at time $t=9.4$. Results are compared with the experimental data by \cite{riboux2010experimental}, for a concentration of $\alpha=1.7\%$, close to that of the numerical configuration. The points have been extracted directly from \cite{risso2018agitation}.}
\label{Fig:pdf3d}
\end{figure} 
We display in figure \ref{Fig:pdf3d} the Probability
Density Function of the vertical $z$ and horizontal $x$ velocities (the $y$ component does not present appreciable statistical differences with respect to the $x$ component). 
The statistics are calculated at the horizontal plane $z=25 d_b$, and are displayed at time $t=9.4$ when all the bubbles have left the plane.
We find the same characteristics reported in experiments ~\citep{riboux2010experimental}, against which results are 
 compared.
 The vertical velocity is strongly skewed, indicating a more important probability of having positive fluctuations, while the horizontal components are symmetric.
Furthermore, both components are non-Gaussian, which is related to the complex features of the bubble-induced agitation.
The agreement between numerical simulations and experiments is globally good.
Yet in the numerical experiment the extreme events tend to be more frequent than in experiments, and exponential tails are found for $\sigma \gtrsim 3$. 
This may be related to the fact that the flow is unsteady, and only spatial averaging is done, or experiments may be undersampling extreme events.

\begin{figure}
\center
{\includegraphics[width=0.525 \textwidth]{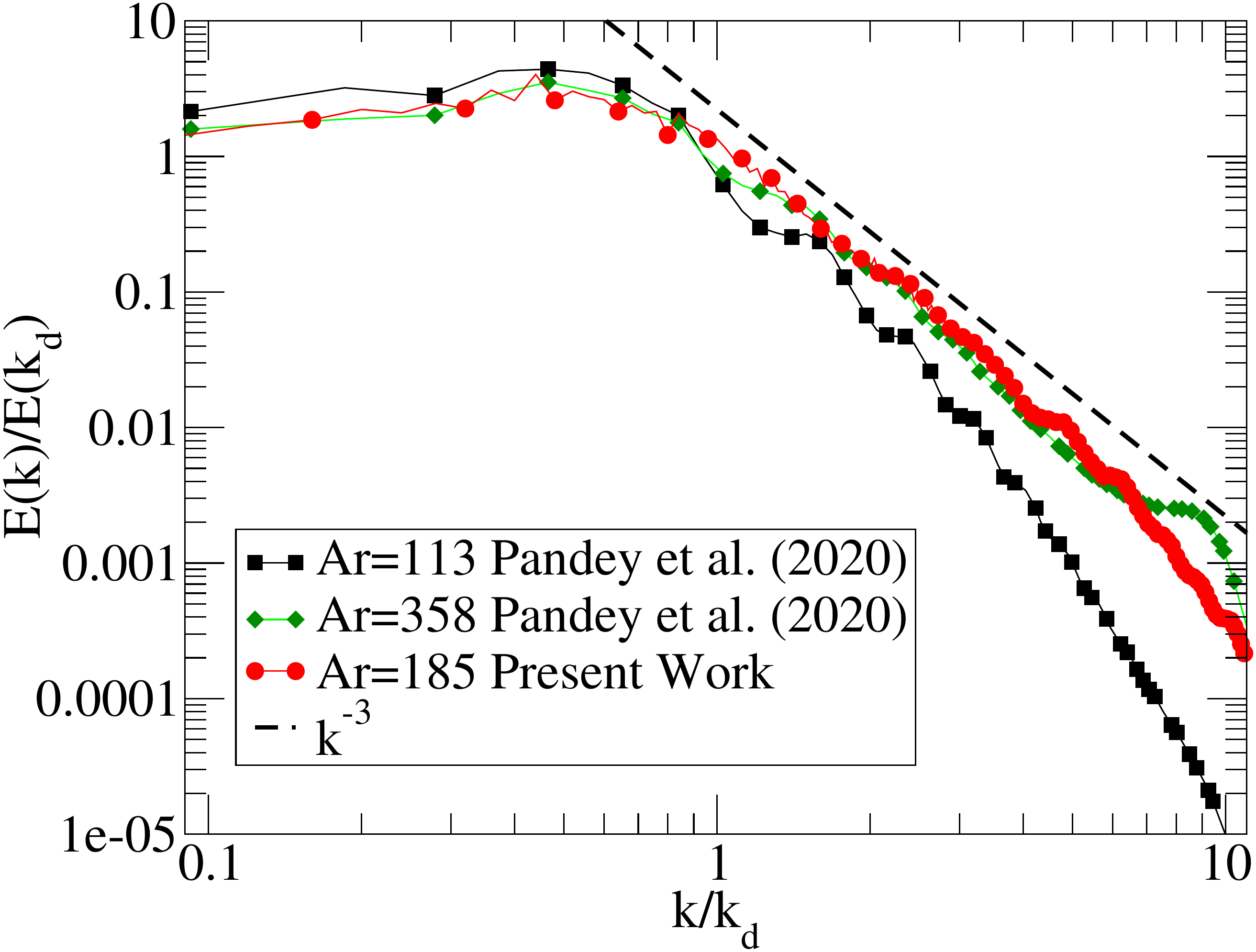}} 
\includegraphics[height=0.425 \textwidth, width=0.425 \textwidth]{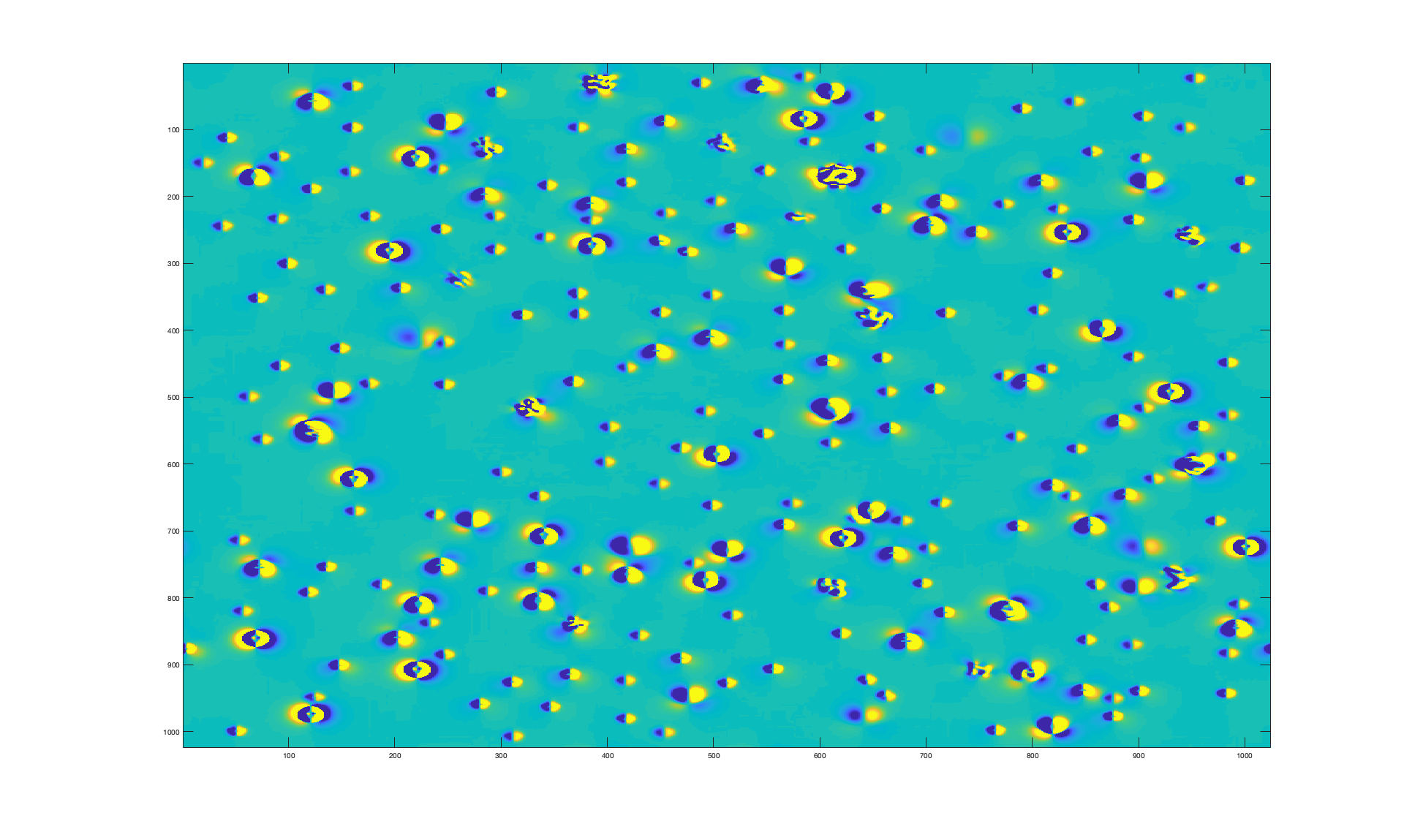}%{\includegraphics[scale=0.5]{fig24b.eps}} \\
%{\includegraphics[scale=0.5]{fig24c.eps}} 
%{\includegraphics[scale=0.5]{fig24d.eps}} \\
\caption{Left panel: Normalised spectrum of the kinetic energy.
Present results ($Ar = 185$ and $Bo = 0.28$) are evaluated at $z=25d_b$ and displayed at $t=9.2$. 
The dashed line represents the $-3$ slope.
%The curve is given to guide the eye.
Data obtained by \citet{pandey2020liquid} are shown for comparison at two different $Ar$ numbers.
Right panel: Vorticity field for bubbles at $t=9$ and $z=25d_b$.}
\label{Fig:spettri3d}
\end{figure} 
In figure \ref{Fig:spettri3d}, we present the spectrum of the kinetic energy computed in the horizontal plane at the middle of the domain $z=25 d_b$.
To compute the spectra, we have interpolated the data on a regular grid. To avoid spurious errors, we have eliminated the highest wave modes, so that spectra are calculated for 512 modes, although the maximum refinement is up to 4096 points.
As for 2D simulations, we have computed the spectrum at different times (not shown here), and we have found very little difference if spectra are calculated around the time when all bubbles have entered the plane used for the computation. When the bubbles have left the spatial region under investigation for a few characteristic times, agitation then decays rapidly, and an exponential fall-off is recorded.
%In particular, we have shown the results at $z=25d_b$, since bubbles were inside at time $t=7.6$, when we start to compute spectra, and left the horizontal plane at about $t=9$. 
%It is clear from the figures that a maximum is reached at around $t=9.2$, before decaying.
%Specifically, the results displayed in figure \ref{Fig:spettri3d} are representative
%In the same way, fields have been found to be non-homogeneous along the vertical direction $z$ even on small distances of about $2 \div 3 ~d_b$, such that we have not performed any average over this direction, 
%Since the horizontal directions have been found to be statistically homogeneous, average on the plane is performed.
Fig. \ref{Fig:spettri3d} shows that bubbles are able to generate significant fluctuations in the length range $\lambda \in [10d_b,0.1 d_b]$, before being dissipated.
After the energy range $\lambda \in [10d_b,1 d_b]$, the spectrum follows a power law with a scaling $E(k) \sim k^{-3}$  in an inertial range over the decade $\lambda \in [d_b,0.1 d_b]$.
%Comparing the dynamics in time of spectra in 2D and 3D, it is apparent that agitation is more persistent in the 2D case.
No hint of an inertial range with slope $k^{-5/3}$ is observed.

For comparison, the very recent numerical results presented by \cite{pandey2020liquid} are also shown for two $Ar$ numbers. 
It is worth recalling, as anticipated in the introduction, that these results have been obtained in coarse-grained simulations (implicit LES) with $24$ points per diameter (instead of $82$ for the present simulations), and using unrealistic physical parameters, notably with a density ratio of about $20$, instead of $800$ as for water/air and used in the present simulations. Despite these important differences, the results obtained in the present DNS are in quite good agreement with the LES obtained in \cite{pandey2020liquid} at $Ar=358$. A departure from DNS only appears at small scales around $1/10 d_b$, because of the lower resolution (256 points used in LES against 4096 used in the DNS here).
The other simulation at $Ar=113$ seems instead to decay faster at all scales.
%\begin{figure}
%\center
%\includegraphics[height=8cm, width=8cm]{fig11.png}
%\caption{Vorticity field for bubbles at $t=9$ and $z=25d_b$. }
%\label{Fig:vort3d}
%\end{figure} 

In order to qualitatively complement this analysis, we show in figure \ref{Fig:spettri3d} (right)  the vorticity field on the same plane used to compute the spectra.
This field highlights the position of the bubbles and  the generation of vorticity at the scale of the diameter and slightly more.
In some cases it is apparent that different vortices have interacted, producing more complex structures.

While energy spectra contain key information about the flow, they cannot be used to disentangle the different mechanisms leading to the observed scalings, and a scale-by-scale analysis can be particularly useful~\citep{alexakis2018cascades}.
For that purpose, we apply a coarse-graining approach developed in a mathematical framework ~\citep{duchon2000inertial,eyink2006onsager} linked to the filtering approach used in Large Eddy Simulations ~\citep{germano1992turbulence}, and recently applied to different turbulent configurations ~\citep{chen2006physical,xiao2009physical,faranda2018computation,dubrulle2019beyond,valori2020weak}. 
More specifically, we have applied this methodology to the velocity field, obtaining informations about the energy flux and the dissipation. 
The advantage with respect to a spectral approach is that one can gain details also on the locality of the cascade, differentiating regions with positive or negative fluxes. 
Moreover, this spatial filtering approach is positive-definite and local in space, and can therefore be applied also in non-homogeneous flows.

In the filtering or coarse-graining approach ~\citep{germano1992turbulence}, the dynamic velocity field $\bu$ is spatially (low-pass) filtered over a scale $\ell$ to obtained a filtered value $\fltr{\bu}_\ell(\bx)$:
\be 
\fltr{\bu}_\ell(\bx) = \int d^3 r\, G_\ell(\br) \bu(\bx+\br) 
\label{eq:filter}
\ee
where $G_\ell$ is a smooth filtering function, spatially localized and such that $G_\ell (\vec r) = \ell^{-3}G(\vec r/\ell)$ where the function $G$ satisfies
$\int d\vec r \ G(\vec r)=1$, and $\int d\vec r \ \vert \vec r \vert ^2 G(\vec r) = \mathcal{O}(1)$. 
By applying the filtering to the Navier-Stokes equations for the liquid phase we obtain the coarse-grained dynamics 
\be 
\partial_{t} \overline{\bu}_\ell 
+  (\overline{\bu}_\ell \cdot \grad)\overline{\bu}_\ell = -\grad\overline{p}_{\ell} 
-\grad\cdot\btau_\ell
+\nu\nabla^{2}\overline{\bu}_\ell
%+{\bf g}.
\label{eq:u-eq-ell} 
\ee
Here  $\btau_\ell$ is the 
{\it subscale stress tensor} (or momentum flux) which describes the force exerted 
on scales larger than $\ell$ by fluctuations at scales smaller than $\ell$. It is given by:
\be 
(\btau_\ell)_{i,j} = 
\overline{(u_i u_j)}_\ell -
(\overline{u}_\ell)_i (\overline{u}_\ell)_j 
\ee 
The corresponding pointwise kinetic energy budget reads
\be
\partial_t \left(\frac{1}{2}|\fu|^2\right) + \partial_j
\widetilde{{ q}}_j
= -\Pi_\ell  
 - \nu|\grad\fu|^2.
\label{kinetic-large}
\ee
where we have dropped the $\ell$ subscript whenever unambiguous for the sake of clarity, and 
\be 
\widetilde{{ q}}_j=\left[ \left(\frac{1}{2}|\fu|^2 +\overline{p} \right)) \overline{u}_j  + \tau_{ij}\overline{u}_i
- \nu\partial_j\left(\frac{1}{2}|\fu|^{2}\right)\right]
~~\text{;}~
\Pi_\ell(\bx) \equiv -(\partial_{j}\overline{u}_{i})\tau_{ij},
\label{kinetic-flux}
\ee
where $\widetilde{{\bf q}}_\ell$ is the transport term, and $\Pi_\ell$ is the sub-grid scale (SGS) energy flux. 
This term is key since it represents the space-local transfer of energy among large and small scales across the scale $\ell$. 
The term $\Pi_\ell$ identifies the presence of a local direct (positive) or inverse (negative) energy cascade according to its sign. 
The last term in Eq. (\ref{kinetic-large}) represents the coarse-grained dissipation $\epsilon_\ell=\nu|\grad\fu|^2$.
If a spatial average is done for different values of the filter width, one can find the average transfer of energy at each scale. 
In this work, we have applied a Gaussian filter defined as:
\begin{equation}
 G ({\bf r}) = \sqrt{\frac{6}{\pi}} \exp(-6 {\bf r}^2),
\end{equation}
as used typically in LES ~\citep{Pope_turbulent}.
Since the flow is homogeneous in the horizontal direction, the filtering can be efficiently performed in spectral Fourier space, multiplying the quantity to be filtered by the Fourier transform of the filter
\begin{equation}
\widehat{G}_\ell ({\bf k}) = \exp(-k^2 \ell^2/24),
\end{equation}
and then transforming back into physical space. 

\begin{figure}
\center
{\includegraphics[scale=0.35]{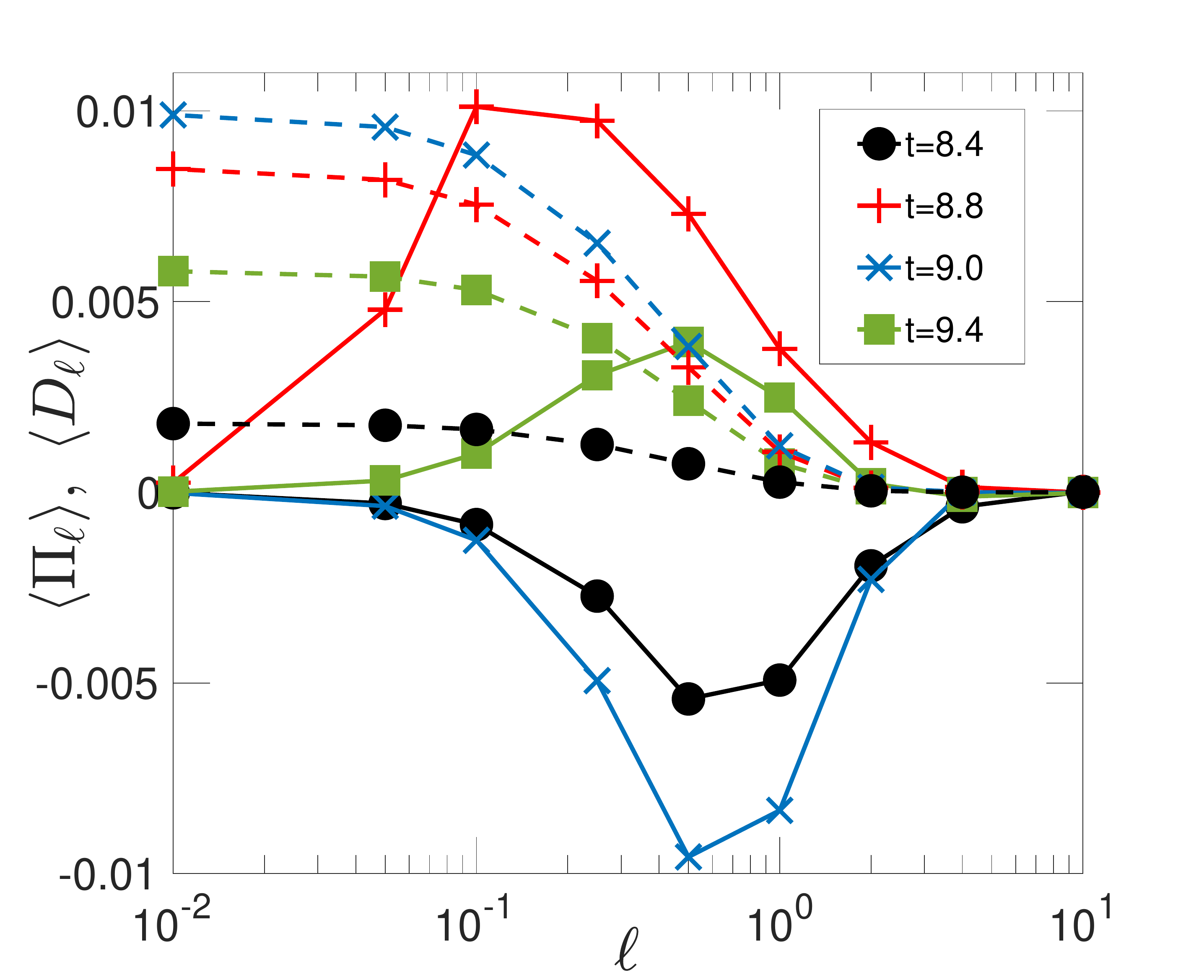}} 
\caption{Mean energy flux at different filter lengths, the energy flux (solid lines) and the dissipative flux (dashed lines) are displayed at different times.
Fluxes are computed at $z=25d_b$, the same as for the computations of spectra. 
The length scale displayed on the $x$ axis is normalised with the diameter $d_b$, so that $\ell=1$ corresponds to the initial bubble diameter.}
\label{Fig:filter3d}
\end{figure} 
In figure \ref{Fig:filter3d}, we show the mean fluxes computed from the coarse-grained quantities defined in Eq. (\ref{kinetic-flux}).
The physical features unfold are the following:
\begin{enumerate}
\item[(i)] The scale-by-scale fluxes show variability in time, pointing out the statistical unsteadiness of the transfer processes. 
\item[(ii)] Both inverse (negative flux) and direct (positive flux) are found. On average, more energy is transferred toward small scales, but the inverse process is not negligible. 
Both cascades involve more than a decade of  scales. The direct cascade is more significant at smaller scales, and the inverse cascade at larger scales.  
\item[(iii)] Energy is transferred from scales around $\ell \approx  d_b$ in both directions.
\item[(iv)] At around the same length scale the dissipation becomes significant.
\item[(v)] At smaller scales dissipation and direct flux become comparable.
\end{enumerate}
Our scale-by-scale analysis points to the following physical picture of the pseudo-turbulent agitation induced by bubbles.
Energy is injected by buoyancy (the only force at play here) and transferred by the bubbles into the liquid 
via the interface at scales comparable to the bubble diameter. The energy input $W_b$ must be proportional to the work made by buoyancy: $W_b \sim \alpha g U_b$.
Dissipation becomes significant at $\ell \sim d_b$, showing that fluctuations are mostly dissipated inside the wakes generated by bubbles.
%Results show that there is roughly a balance between the dissipation and the energy flux.
At smaller scales than the diameter, there is a range where $W_\mathrm{diss} \approx W_b$, which means  $\nu (\delta u_\ell)^2\ell^2 \sim  \alpha g U_b $, where we have considered the two-point quantities $\delta u_\ell=u(x+\ell)-u(x)$.
 This gives the scaling behaviour $\delta u_\ell^2 \sim \ell^2 $, which means in spectral space $E(k) \sim k^{-3} $. 
The argument is similar to what was proposed in previous experimental works~\citep{lance1991turbulence,prakash2016energy}.
%The inverse cascade indicates the formation of the  wakes, which are found to develop up to some characteristic lengths.
%Because of the high Reynolds number, inside the wakes smaller scales are formed and a cascade is triggered, which is linked to the direct cascade. 
%
%
%While the scaling behaviour appears to be fixed on dimensional grounds by the balance in turbulent fluxes, the result shows that the field caused by bubble agitation is smooth at large wavenumbers, consistently with the wake model proposed by ~\citep{risso2011theoretical}. 

\begin{figure}
\center
\includegraphics[scale=0.33]{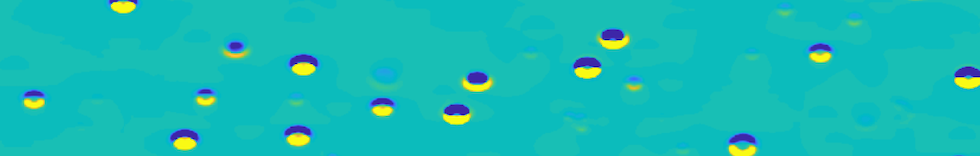}
\includegraphics[scale=0.33]{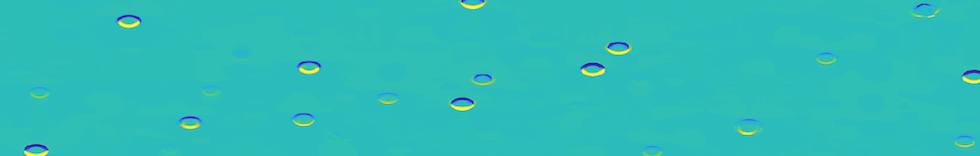}
\caption{Local energy flux at $t=9$ and $z=25d_b$. In the upper panel the filter length is $l=0.5 d_b$, in the lower panel $l=0.1 d_b$. The color scale is the same in both panels. $1024$ points are taken in the horizontal direction, and $256$ in the vertical direction.}
\label{Fig:flux3d}
\end{figure} 
To further understand the mechanisms indicated, we show in figure \ref{Fig:flux3d}  a slice of the energy flux at two different scales: the diameter and a smaller scale.
The pictures show that the energy flux and hence dissipation are concentrated in the wakes generated by the bubbles.
Furthermore these structures, initially at the scale of the diameter, may become a little larger, indicating the generation of larger eddies, and are eventually dissipated at small scales, where the imprint of the bubbles is still detectable.

\section{Conclusions}
\label{sec:conc}

%
%We first assess the influence of the density ratio between phases and the grid refinement using some simpler configurations which can be analysed with one bubble, as done recently by \citet{loisy2017buoyancy}.
%
%
%%In accordance with previous studies~\citep{cano2016paths}, we have found that in order to reproduce the correct bubble dynamics it is necessary to use a realistic density ratio (more than $100$), and a resolution that increases proportional with the bubble Reynolds number.
%
%
%On the basis of the 2D results, we have built up the 3D numerical experiment, which fulfils the requirements in terms of realistic physical parameters and numerical resolution to be considered the first direct numerical simulation (DNS) of such a bubbly flow.
%
%In this configuration, we have analysed the one-point probability density function of the velocity fluctuations and the two-point statistics looking at the Energy spectra.
%
We have numerically investigated buoyancy-driven bubbly flows, focusing on the agitation induced by the bubbles on the fluid. The purpose of the study was to characterise the physics of the collective motion induced when many bubbles rise under the sole effect of gravity. We have carried out several 2D and 3D preliminary tests and the first high-resolution direct numerical simulation of a 3D realistic flow.

We have first extensively investigated 
the interplay between the numerics and the physics of bubbly flows at moderate and high Reynolds numbers in order to properly set numerical parameters compatible with a reliable description of the flow. To do so, we studied different configurations and compared the results with recent studies made using different interface advection methods. 
These numerical experiments have shown on the one hand that the physical parameters, and most notably the density ratio of the two fluids, may affect the results both qualitatively and quantitatively.
On the other hand, to be sure to have solution at convergence, the spatial resolution should be increased when increasing the $Re$ number.
In particular, to carry out a DNS it seems necessary to fulfil the following criteria:
(i) The density ratio has to be realistically high $\rho_l/\rho_b > 100$ ; (ii) The viscosity ratio should also be realistic $\mu_l/\mu_b \approx 100$ ; (iii) The number of points used to resolve the bubbles must increase linearly with the Archimedes number (or the Reynolds number based on the raising velocity). As a rule of thumb, this number should be of the order $ d_b/\Delta \approx {Ar}/2$.
A numerical simulation which fails to fulfill these criteria cannot be called a DNS but should be considered as an implicit Large-Eddy-Simulation without sub-grid modelling. 

Given the numerical constraints which do not allow a parametric DNS study of a high-Reynolds flow in three dimensions, we have  rather performed a comprehensive analysis of the liquid agitation in a two-dimensional bubble column at moderate and high Reynolds numbers with a volume fraction of about $5\%$ in the bubble layer, in order to prepare a three-dimensional study. 
Both unsteady and steady numerical experiments have been carried out.
In this configuration, we have analyzed the velocity fluctuations in both phases and find different behaviours for different Reynolds numbers, even if the $-3$ slope of the spectra seems to be a robust feature of this type of flow, also in 2D. 
In particular the presence of an inverse cascade at scales larger than the diameter has been unveiled for flows at $Ar$ number higher than $100$.
Indeed, at larger scales than the diameter, where the dissipation is negligible, the energy budgets is $\Pi_\ell \sim W_b \approx \alpha g U_b$ which gives the Kolmogorov scaling $\delta u_\ell^2\sim \ell^{2/3}$, or $E(k) \sim k^{-5/3} $, typical of an inverse cascade.
As expected, PDFs show a strong anisotropy of the fluctuations in the vertical direction, while horizontal fluctuations are symmetric. 
The 2D simulations have indicated that the statistics obtained in unsteady simulations are accurate, provided the time and space windows used to keep the data are well chosen. We have provided all the criteria to be fulfilled to get a reliable numerical experiment. 
Besides the main numerical relevance, the configuration may have some similarity to  that investigated experimentally in a confined two dimensional configuration ~\citep{bouche2012homogeneous,bouche2014homogeneous}.

Then, on the basis of the results obtained in the previous tests, we have performed a single numerical experiment of  a 3D bubble column at $Ar=185$, which corresponds to a Reynolds number consistent with experiments.
First we have observed that the one-point PDF of the velocity of the liquid agitation in numerical simulations   are in agreement with those obtained experimentally ~\citep{riboux2010experimental,risso2018agitation}. 
However the tails related to rare events ($>3 \sigma$ from the mean) are more pronounced, with an exponential decay, than in the experiments where they are more Gaussian.
It is difficult to say if this discrepancy is due to the strong unsteadiness of the flow, or to a smoothing of extreme events in real experiments because of the presence of impurities.

The energy spectra have also been analysed and compared with recent numerical simulations performed at low resolution and low density ratio.
We have found a $k^{-3}$ scaling over a decade of scales smaller than the diameter, and possibly at scales just a little larger. We have not found any hint of a Kolmogorov $k^{-5/3}$ scaling neither at large or small scales.

We have shown through a scale-by-scale analysis in physical space that the spectra are related to a nonlinear cascade mechanism, and do not reflect only the presence of wakes. 
Indeed we have found that a flux of turbulent kinetic energy is present in the range of scales going from $2d_b$ up to $d_b/20$, where dissipation becomes dominant.
In this range the balance between the flux of energy and the dissipation explain the  $k^{-3}$ scaling.
Interestingly our unsteady numerical experiment highlights the presence of instantaneous fluxes in both directions indicating the tendency to create locally larger structures around the bubbles, even though on average the energy is injected around the bubble diameter scale and mostly transferred to smaller scales where it is eventually dissipated.

An important result of our work comes also from the comparison with the recent simulations by  \cite{pandey2020liquid}.
According to our analysis these simulations should be considered as \emph{implicit} LES when $Ar>50$, given the low resolution with respect to the bubble size, yet they are representative of the numerical resolution used in most of the works presently carried out in turbulent bubbly flows \citep{elghobashi2019direct,cifani2020flow}.
The present DNS results show that one-point and two-point statistics are in good agreement with the LES obtained at high $Ar$ number, except at small scales.
The present conclusion is hence that this kind of LES, using only $20-30$ points to resolve the bubble diameter, seems to be sufficient to get consistent results with respect to large-scale statistics, although finite Reynolds number effects are found to be exaggerated.
At variance with what was found for single-bubble observables~\citep{cano2016paths}, our results validate the use of LES to analyse large-scale collective properties in turbulent bubbly flows.

Concerning future developments, it would be interesting to analyse the budgets of the momentum and energy equation in relation to the development of two-fluid models, that is Reynolds-Averaged Navier-Stokes (RANS)~\citep{Dre_83,drew2006theory}.
It has been known for a long time that numerical stability is not assured in two-fluid models~\citep{stuhmiller1977influence,ramshaw1978characteristics}.
% In such a case, the equations are not hyperbolic and the discretized equations do not allow a grid-converged solution to be achieved. Unstable modes in the solution appear, severely affecting the model prediction and its sensitivity to grid refinement. 
    To ensure the hyperbolicity of the system and therefore its stability, an ad-hoc pressure term is usually introduced by hand ~\citep{prosperetti1987linear,tiselj1997modelling,davidson1990numerical,song2001one}.
   While the functional dependence on the volume fraction seems to be clearly established ~\citep{panicker2018hyperbolicity}, 
   the multiplicative factors are unknown and may be found only from DNS or experiments.  

\section{Acknowledgements}
This work was granted access to the HPC resources of [TGCC/CINES/IDRIS] under the allocation 2019- [A0062B10759] attributed by GENCI (Grand Equipement National de Calcul Intensif).
We thank Rodney Fox for fruitful discussions. 

%\clearpage
%\bibliographystyle{unsrt}
\bibliographystyle{jfm}
\bibliography{paper}

\clearpage
\appendix 

\section{Arrays of bubbles}
\label{app:array}

For the test case proposed by ~\cite{esmaeeli1999direct}, displayed in Fig. \ref{Fig:3D-tryg},
the details of the different grids are reported in table \ref{tab:grid3D}, together with the steady values of the Reynolds number.
\begin{table}
\begin{center}
\begin{tabular}{ l | c c c c }
\hline
%\toprule
N & $16^3$ & $32^3$ & $64^3$ & $128^3$ \\
\hline
$d_b / \Delta$ & $10$ & $20$ & $40$ & $80$ \\
\hline
$Re_f$ (present)  & $23.05$ & $22.01$ & $21.275$ & $21$ \\
\hline
$Re_f$ ~~\citep{esmaeeli1998direct} & \textbackslash & \textbackslash & $20.49$ & \textbackslash \\
\hline
$Re_f$ ~~\citep{loisy2017buoyancy} & $19.05$ & $20.22$ & $20.58$ & \textbackslash \\
\hline
%\bottomrule
\end{tabular}
\caption{Grid resolutions and final Reynolds number for the 3D array of bubbles of ~~\cite{esmaeeli1999direct}.}
\label{tab:grid3D}
\end{center}
\end{table}
%\subsection{3-D Stokes flow}

%\subsection{3-D array at moderate Reynolds number}

%\subsection{3-D oblique rise of bubbles}

\begin{figure}
\center
{\includegraphics[width=.4\textwidth]{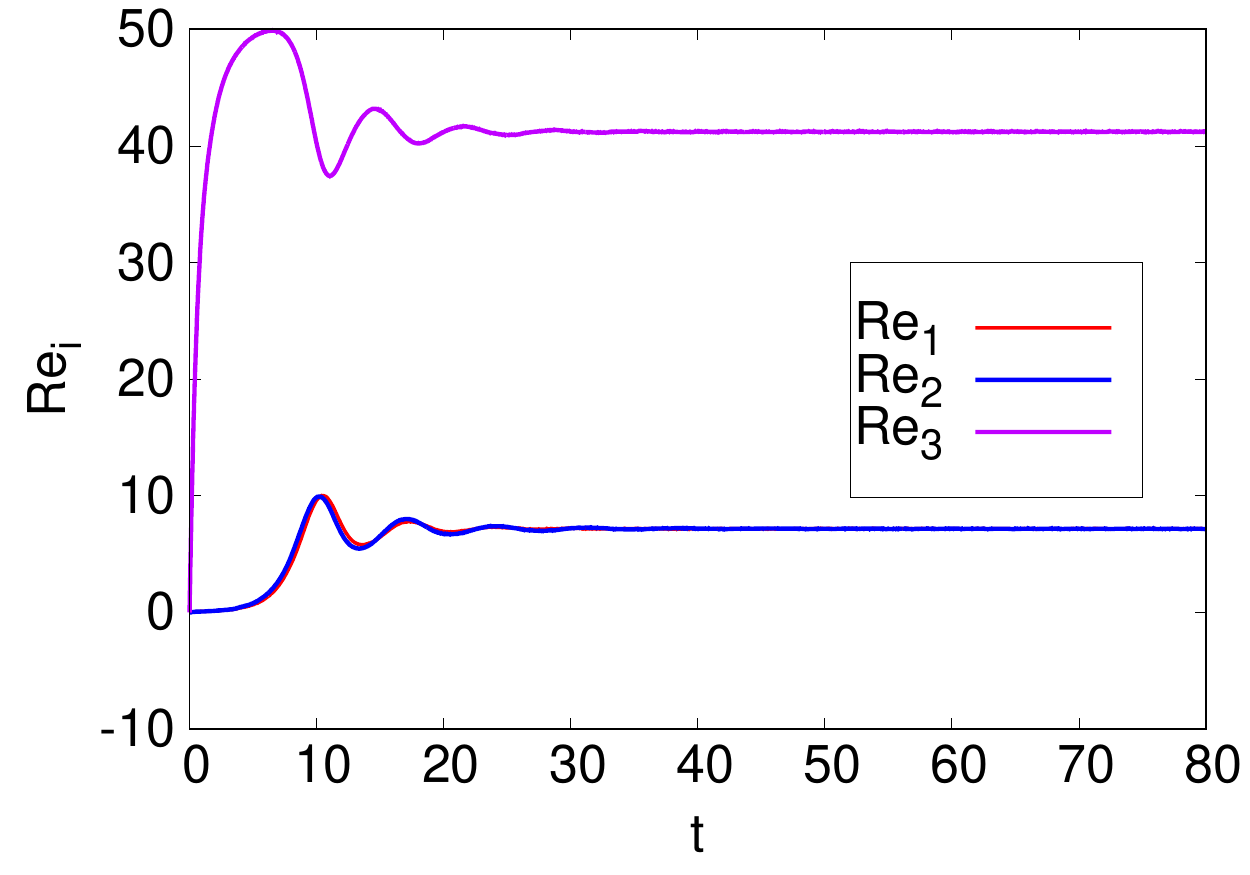}}  \hspace{1cm}
{\includegraphics[width=.4\textwidth]{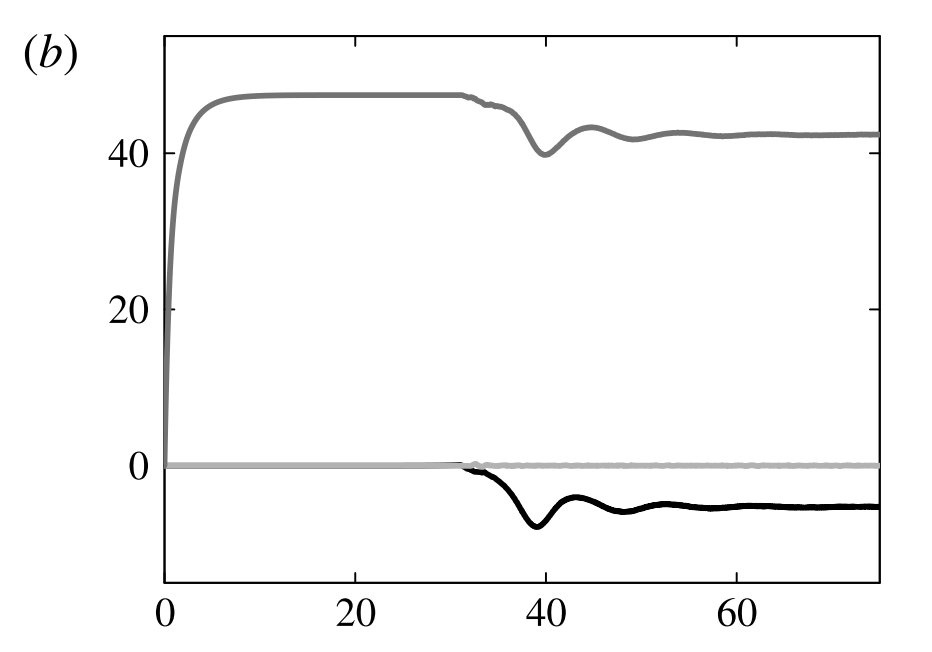}} \\
{\includegraphics[width=.4\textwidth]{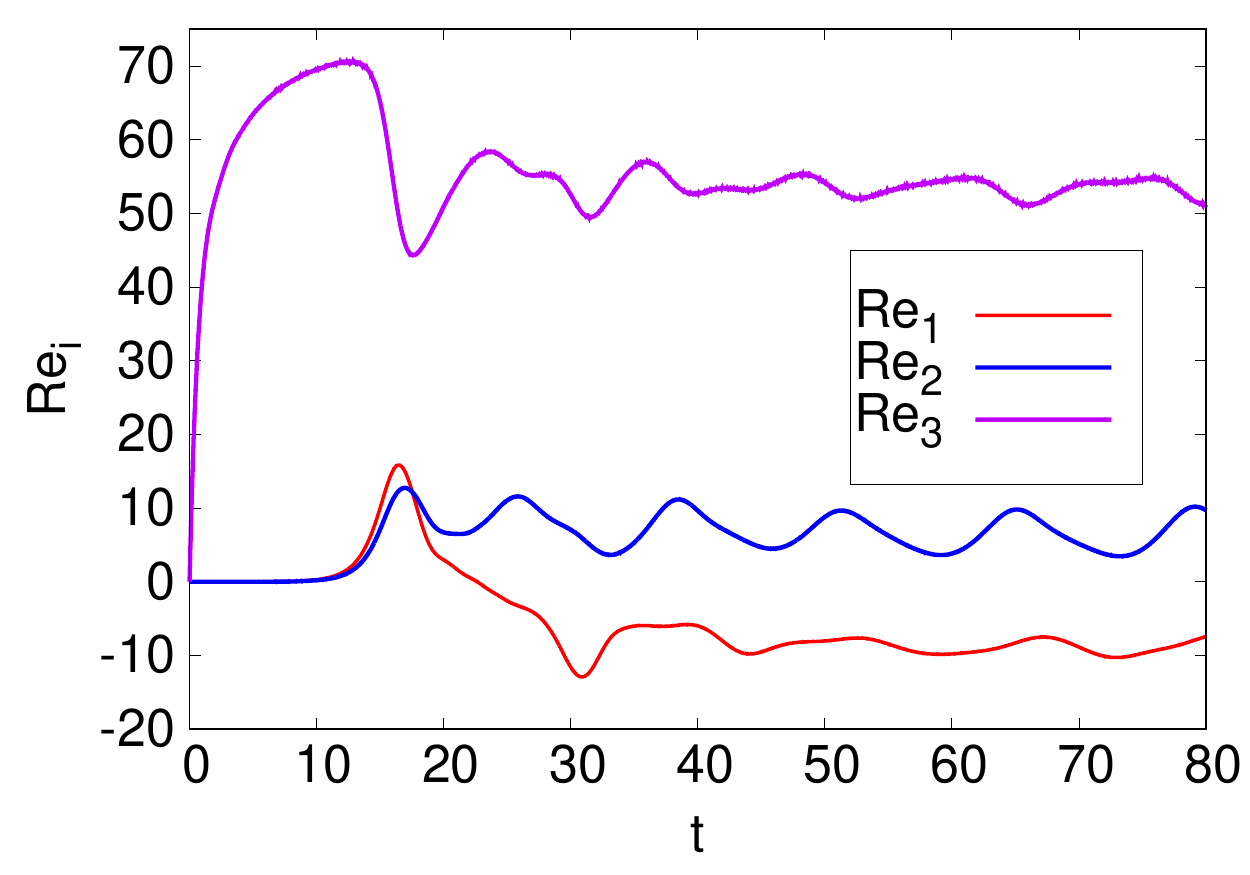}}  \hspace{1.cm}
{\includegraphics[width=.4\textwidth]{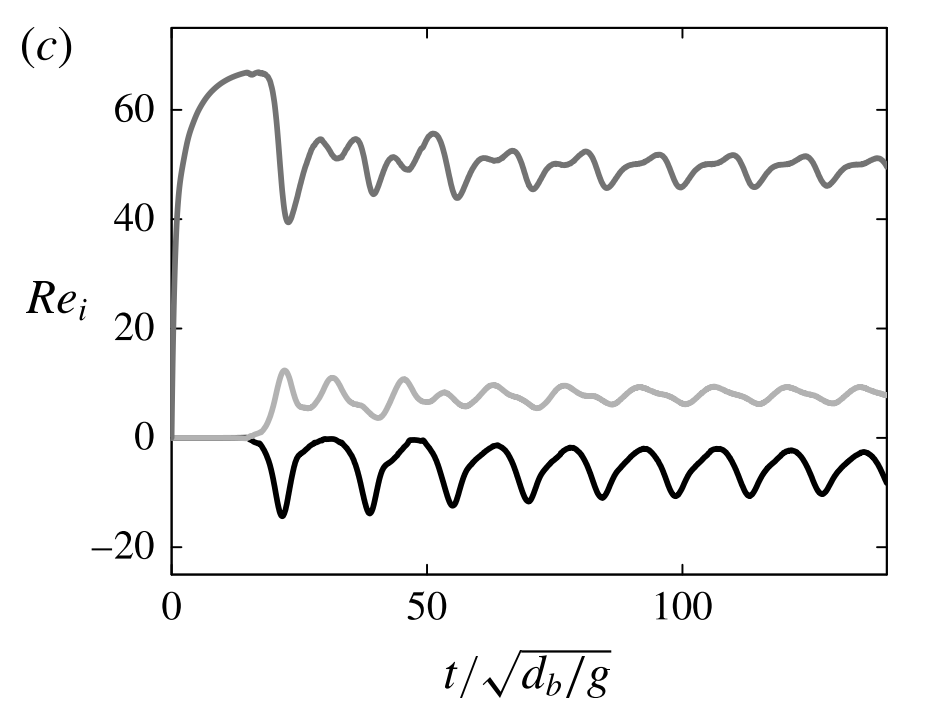}} \\
\caption{Time evolution of two components of the bubble Reynolds number for regime (b) (top panels) and (c) (bottom panels). 
Present results on the left panel, DNS by~~\citep{loisy2017buoyancy} on the right panel.}
\label{Fig:3Doblique-c}
\end{figure}

\begin{table}
\begin{center}
\begin{tabular}{ c  c  c  c}
\hline
%\toprule
Case & $Ar$ & $Bo$ & $\phi$ \\
\hline
%\midrule
$a$ & $29.9$ & $2$ & $0.008$ \\
$b$ & $40.7$ & $0.38$ & $0.13$ \\
$c$ & $40.7$ & $0.38$ & $0.038$ \\
\hline
%\bottomrule
\end{tabular}
\caption{Non-dimensional parameters for the 3D-oblique test case.}
\label{tab:oblique}
\end{center}
\end{table}
The parameters of the simulations of the case of the oblique array of bubbles are reported in table \ref{tab:oblique}. 
The results not shown in the main text are given in figure \ref{Fig:3Doblique-c}, and summarised in table \ref{tab:oblique2}.
Similar regimes are captured in each case, while the transition may occur at different times compared to ~\cite{loisy2017buoyancy}, since it is triggered by numerical asymmetry. 
For the same reason, while we expect a quantitative agreement in the direction of gravity, the other two components can share the energy in a different way, provided that this is compatible with the symmetry of the problem.
The steady value of the different components of the bubble Reynolds number is in excellent agreement for cases (a) and (b), while in case (c) where a steady regime is not reached, we can quantify the accuracy by comparing the oscillation periods, which are found to be in close agreement.
\begin{table}
\begin{center}
\begin{tabular}{ l  c  c  c}
\hline
Case & $a$ & $b$ & $c$  \\
\hline
$Re_1$ (present)  & $34.1$ & $41.2$ & $52$  \\
$Re_1$ ~~\citep{loisy2017buoyancy} & $34.4$ & $42.2$ & $50.1$\\
\hline
$Re_3$ (present)  & $1.41$ & $7.15$ & $-8.5$  \\
$Re_3$ ~~\citep{loisy2017buoyancy} & $1.4$ & $-4.15$ & $-8.1$\\ %\midrule
\hline
%\bottomrule
\end{tabular}
\caption{Final $Re$ for the 3D-oblique test case. For the last oscillating case we have reported the average over the last steps, so the comparison is to be considered qualitative.}
\label{tab:oblique2}
\end{center}
\end{table}
\section{Technical issues}
\label{app:tech}
 
\subsection{Grid refinement effects}
%\begin{figure}
%{\includegraphics[scale=.15,angle=90]{fig7a}} 
%{\includegraphics[scale=.15,angle=90]{fig7b}} 
%{\includegraphics[scale=.15,angle=90]{fig7c}} 
%\caption{The vorticity field is displayed together with the mesh for three choices of the grid resolution, namely we have changed the error threshold as $e_v = 0.001; 0.003; 0.01$, in absolute value.
%%\color{red} SP: Need legend, The figures should use a color field to display resolution, rather than the grid itself.
%%
%}
%\label{Fig:bolla1}
%\end{figure} 
\begin{figure}
{\includegraphics[scale=0.5]{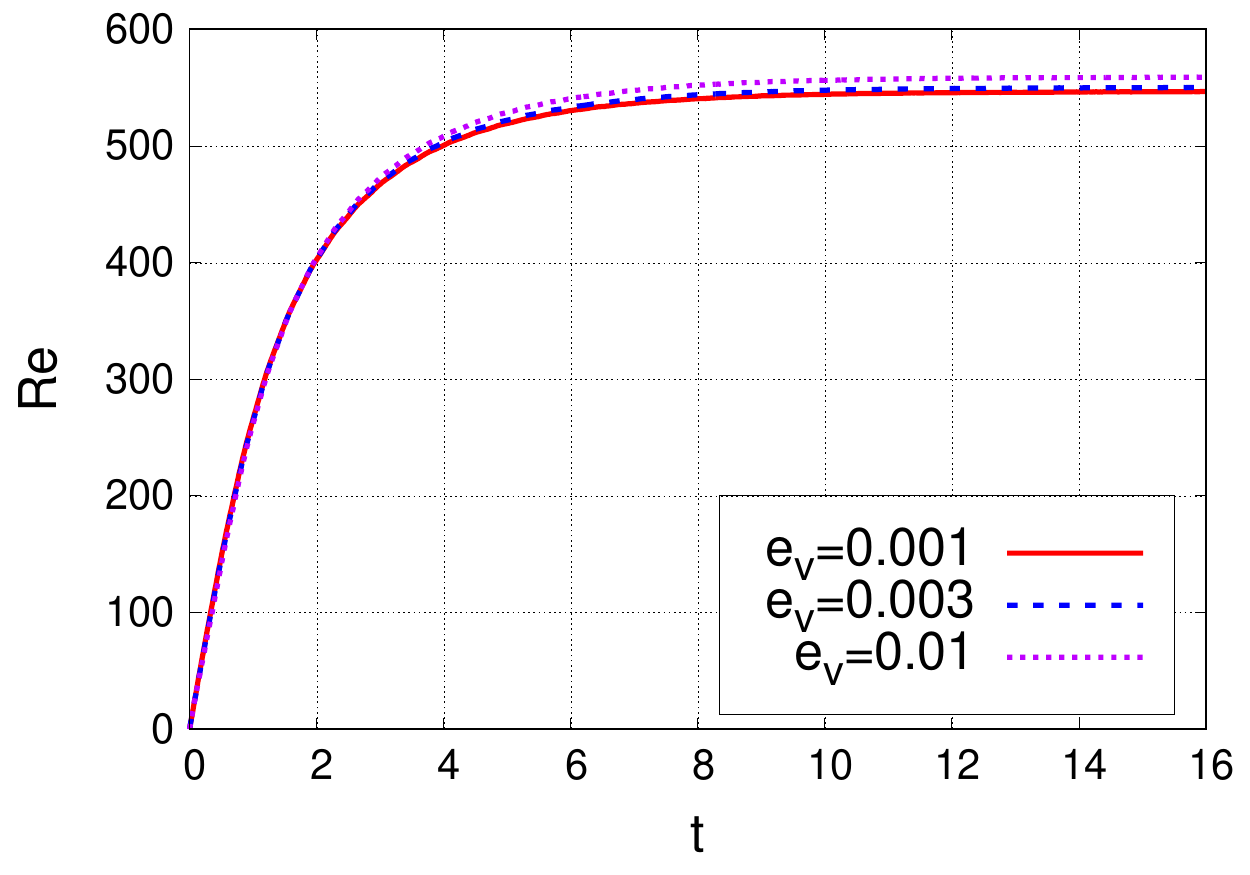}} 
{\includegraphics[scale=0.5]{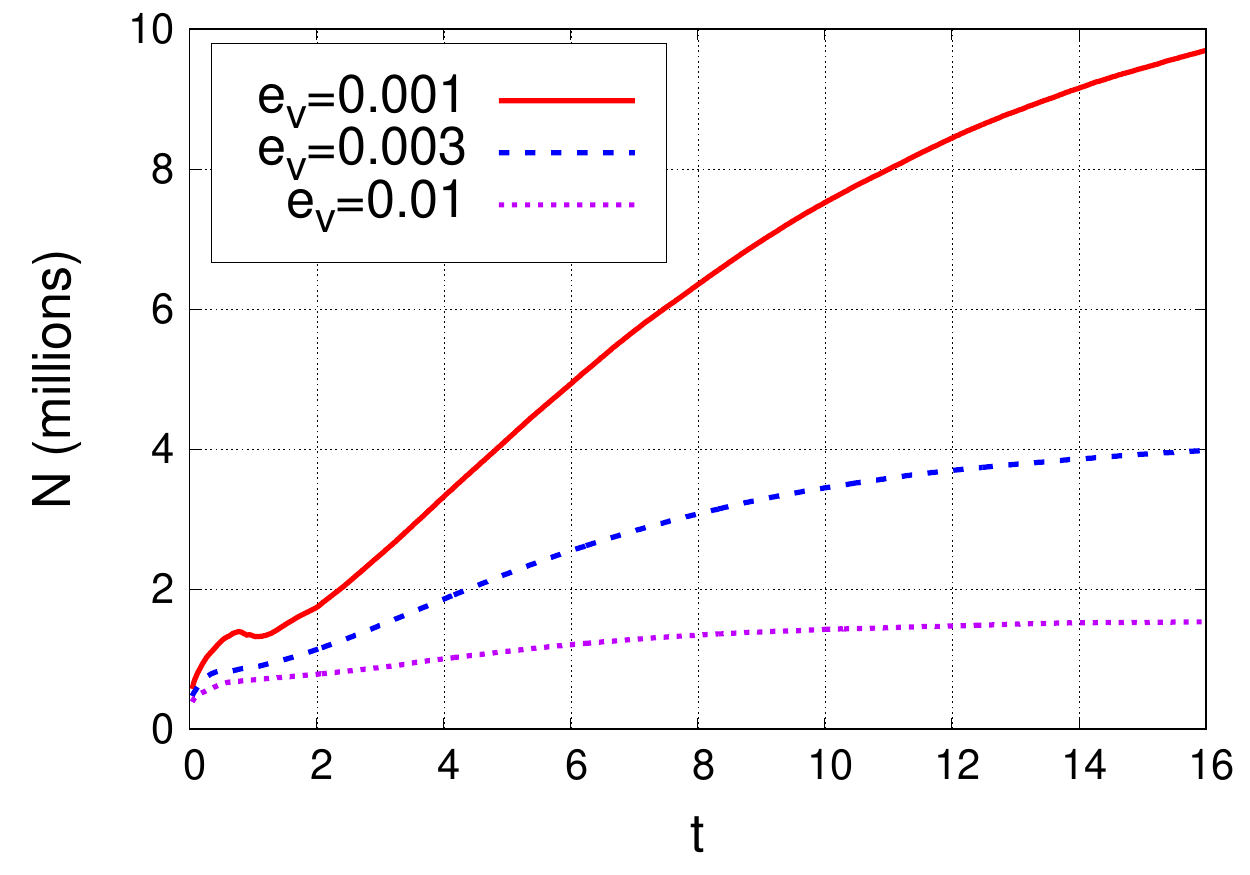}} 
\caption{(a) The average Reynolds are plotted for the three resolutions. The final values are: $Re=559$ for $e_v=0.01$, $Re=549$ for $e_v=0.003$ and $Re=547$ for $e_v=0.001$.
(b) The number of mesh points is plotted against time. It should be noted that the number of points is approximately steady for the two coarser resolutions, while it is growing rapidly for the most refined resolution.}
\label{Fig:bolla1b}
\end{figure} 
We have replicated some of the results obtained by ~\cite{cano2016paths} to study the behaviour of a single bubble rising ``in a large tank'' i.e. far from any boundaries.
We have used only the physical parameters used for the realistic 3D bubble column. 
Namely, we fix $Ar =185$ and $Bo = 0.28$.
%We choose as length unit the diameter of the bubble. The domain is $120^3$.
The acceleration of gravity is set to unity, which gives a characteristic rise velocity also of order unity, and a maximum time for the simulation comparable to the domain size.
In this regime, it turns out that bubble trajectories are between the rectilinear and chaotic regimes, as found in the original paper ~\citep{cano2016paths}.
We have simulated the bubble rise with  three different grids,
namely varying the threshold of the error tolerance, fixed at $err_v=0.001; 0.003; 0.01$, in absolute value.
This threshold controls the local refinement of the grid~\citep{van2018towards}.
In Fig. \ref{Fig:bolla1b}a, we show the evolution of the rise velocity of the bubble, which is given by the Reynolds number in dimensionless form.
The two low-tolerance grids show very little difference (less than $1\%$) in the rise velocity, whereas for the highest-tolerance grid the difference in rise velocity is of the order of $5\%$.
This indicates that the three grids are sufficient to get a qualitative reproduction of the physics of the problem but that only the two more refined are at convergence.
In figure \ref{Fig:bolla1b}b, we display the evolution of the number of grid points with time for the three different grids. This gives a measure of the computational cost of each setup.
From figure \ref{Fig:bolla1b}a, we can see that a transient is present with a duration of about $8\div10$ unit times. 
Results show that an over-refinement of the bubble is present for the lowest error threshold.
We have therefore found that convergence is reached with $N_{Max}=2^{12}$, such that the maximum refinement is of $82$ points per diameter, with an error threshold of $0.003$ (absolute value) in the velocity.  
This resolution has hence been chosen for the final 3D bubble column simulation.
 
\subsection{Coalescence}
\begin{figure}
\begin{center}
{\includegraphics[scale=.15]{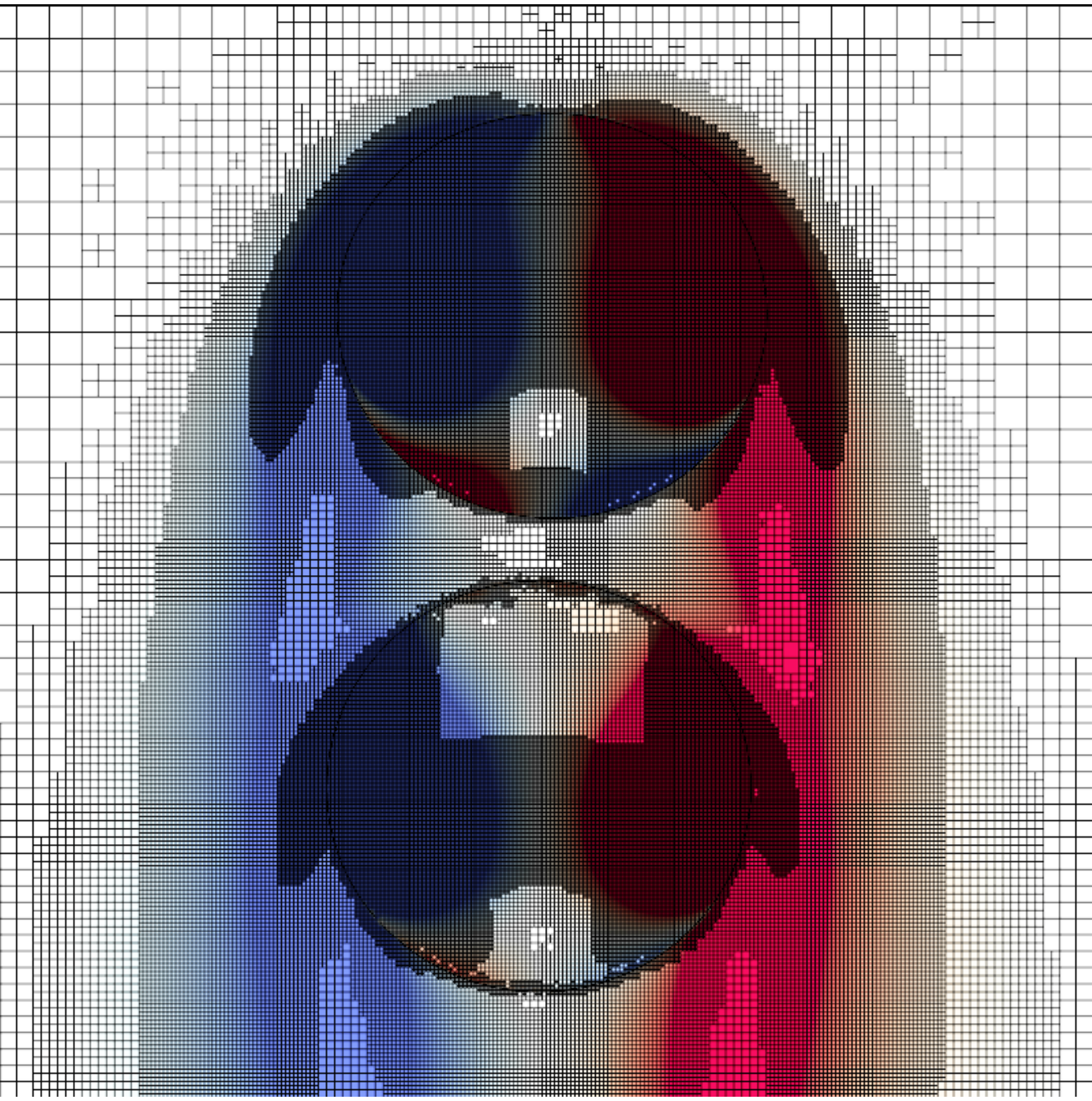} 
\includegraphics[scale=.15]{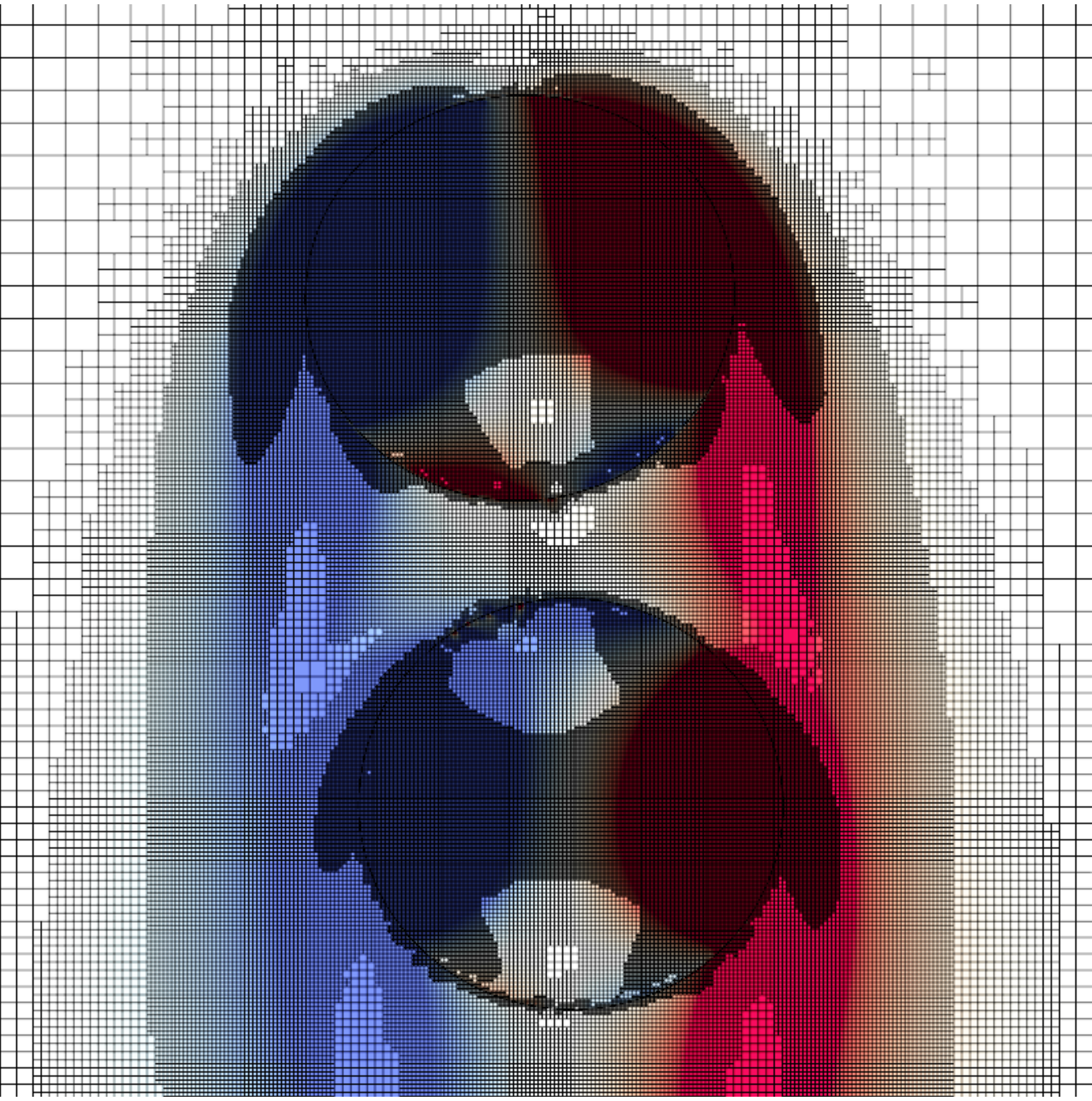}\\
%\includegraphics[scale=.15]{./coalescenza/L12rho100/bubble_0844.pdf}
%\includegraphics[scale=.15]{./coalescenza/L12/bubble_0770.pdf}
% \\
\includegraphics[scale=.15]{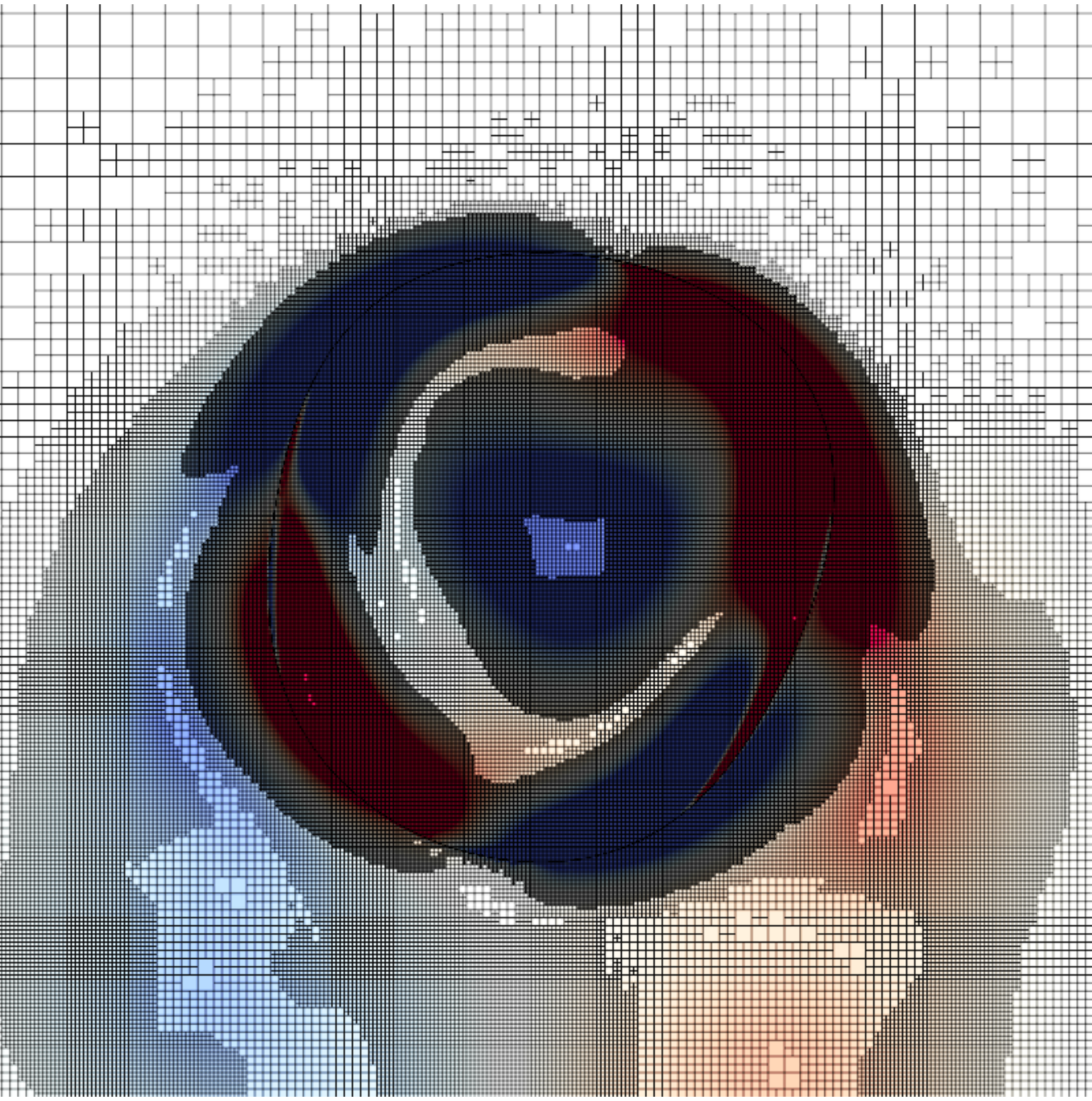}
\includegraphics[scale=.15]{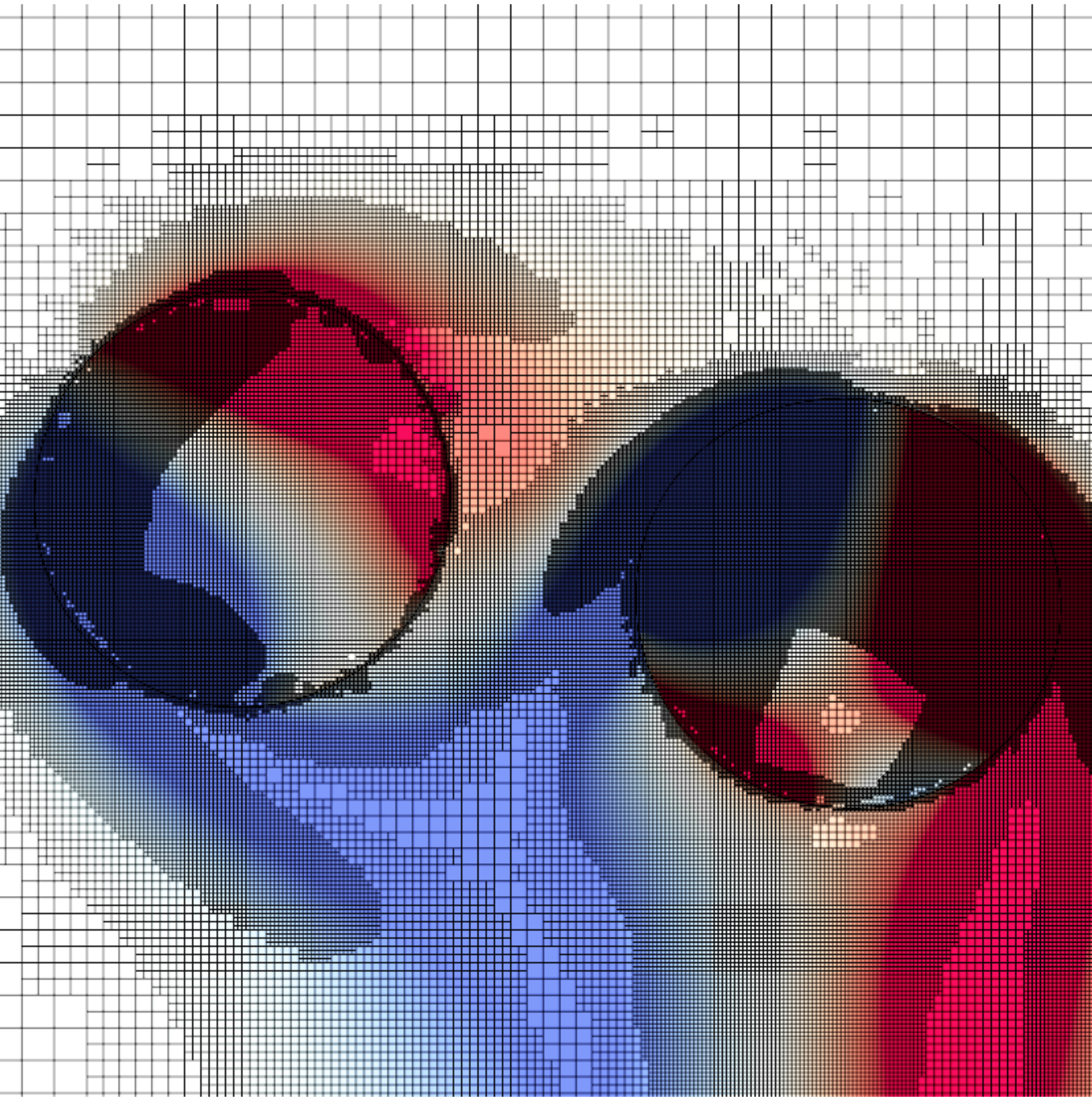}
\\}
\caption{Contour plot of the vorticity at two instants. The grid is drawn to show the degree of refinement obtained with $N_\text{Max}=2^{13}$. Left column: simulation with $\rho_l/\rho_b=100$; Right column: simulation with $\rho_l/\rho_b=1000$.  } 
\label{Fig:coa1}
\end{center} 
\end{figure} 
We have studied from a qualitative point of view the coalescence of two bubbles in relation to density ratio and grid refinement.
This is a vast area of research ~\citep{liao2010literature} and a detailed analysis of this issue is out of the scope of the present work. 
However it is important to have some control on this process to avoid spurious effects. 
%It also constitutes an interesting configuration to test the influence of the numerical setup.
Generally speaking, present attempts to carry out DNS of bubbly flows may use methods which prevent coalescence entirely ~\citep{tryggvasondirect,loisy2017buoyancy}.
%This is convenient from a practical  point of view, but cannot be considered entirely satisfactory since in general turbulent flows, in presence of buoyancy, coalescence may happen. 
On the contrary, VOF methods, like that used in the present work, tend to make coalescence too easy ~\citep{scardovelli1999direct}, if numerical parameters are not well controlled. 
  In a recent work, coalescence has been observed 
  in VOF simulations of turbulent droplet-laden flows ~\citep{dodd2016interaction}, but the issue of a possible spurious impact of the numerics has not been considered.
Here we consider for this purpose two bubbles in a two-dimensional box of side $20$ times the diameter of the bubbles with periodic boundary conditions.
 The physical parameters are fixed in such a way that dimensionless numbers are $Ar=30$, $Bo=0.1$ and $\mu_b/\mu_l=100$.
%Unfortunately, due to the limited understanding of the  processes underlying coalescence, empirical correlations are still needed. To date, no satisfactory models taking into account all mechanisms and applicable to a wide range of conditions are available in the literature ~\citep{jakobsen2005modeling}.
Based on physical intuition and  empirical evidence, the rate of coalescence depends on the diameters of the bubbles, their relative velocity and the turbulent rate of energy dissipation in turbulent flows.
We consider two bubbles, one on top of the other, initially at rest in a quiescent fluid. The top bubble is at $0.75$ diameter from the bottom bubble.
They start moving because of buoyancy which induces vorticity fluctuations and creates wakes.
%From a physical point of view, those are the mechanisms which bring bubbles together and, with a certain probability, make them coalesce.
With the physical parameters and initial conditions chosen, the probability of coalescence is not zero but quite low, on the basis of standard empirical models ~\citep{prince1990bubble}.
Hence we shall consider our numerical approach satisfying if coalescence is avoided in this case.
We have carried out the simulation of the same test case with the physical sound density ratio $\rho_l/\rho_b=1000$, with different grids.
We have found that convergence is attained with  $N_\text{Max}=2^{12}$, since the results are the same as those obtained with $N_\text{Max}=2^{13}$. 
Using $N_\text{Max}=2^{11}$ instead the coalescence occurs (results not shown here).

Two instants for the maximal resolution are displayed in the top of the figure \ref{Fig:coa1}.
The other resolution are not shown for the sake of simplicity.
Then, we have assessed the influence of the density ratio.
We have chosen the finest resolution  $N_\text{Max}=2^{13}$ to be sure to avoid any discretisation effect.
In figure \ref{Fig:coa1}, we show two instants of this dynamics, displaying also the vorticity field, for two different density ratios at a resolution of $N_\text{Max}=2^{13}$.
Notably in the left column of figure \ref{Fig:coa1} we display the results obtained for a density ratio of $\rho_l/\rho_b=100$, for which coalescence happens.
The coalescence involves several processes as expected by the theory: (1) the two bubbles collide, trapping a small amount of liquid between them; (2) bubbles keep in contact till the liquid film drains out to a critical thickness; (3) the film ruptures resulting in coalescence. Our simulations seem to indicate that in this particular case, dominated by buoyancy the coalescence happens following the classical drainage model ~\citep{shinnar1960statistical}.
Indeed turbulent fluctuations and relative velocity are not sufficient to trigger a faster process.
In the right column, we show the same case but with a density ratio of $\rho_l/\rho_b=1000$.
In this case, coalescence does not occur.
We have investigated different density ratios in the range $\rho_l/\rho_b\in [10,1000]$ (not shown here for the sake of clarity), and it turns out that the threshold for avoiding the coalescence is about $\rho_l/\rho_b=200$.
%Our results therefore point out that to reproduce multiphase flows, a realistic density ratio is mandatory, while it possibly may be slightly less than the actual one.
%\begin{figure}
%{\includegraphics[scale=.21]{./fig6a.pdf}} 
%{\includegraphics[scale=.21]{fig6b.pdf}} 
%{\includegraphics[scale=.21]{fig6c.pdf}} 
%\caption{Contour plot of the vorticity at three instants. The grid is drawn to show the degree of refinement. It is possible to see that even with the present grid which is well refined ($N_\text{Max}=2^{11}$) some details of the wake and of the film surrounding the bubbles are lost with respect to the more refined simulations displayed in Fig. \ref{Fig:coa1} right column.}
%\label{Fig:coa2}
%\end{figure} 

\section{2D induced turbulence: numerical complements}
\label{app:C}

\begin{figure}
{\includegraphics[scale=0.5]{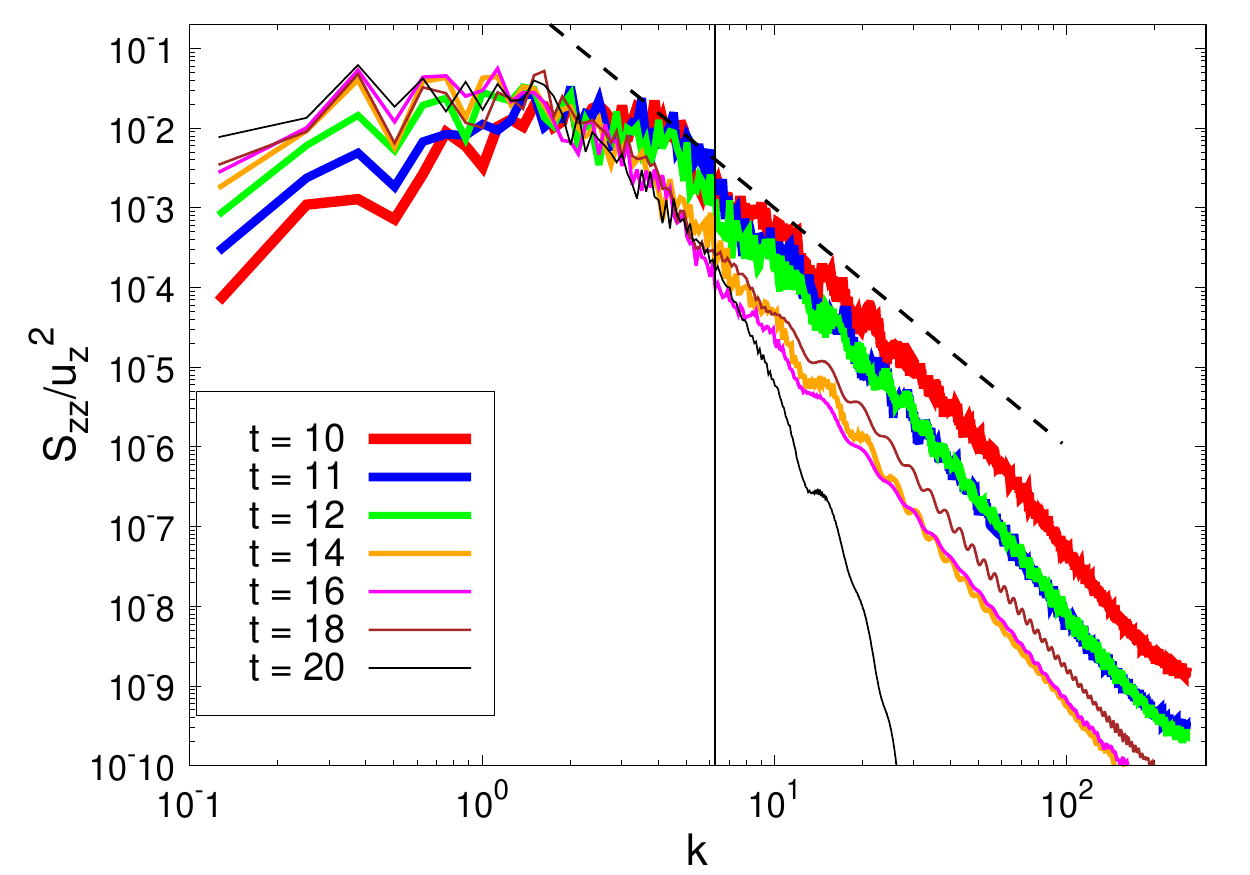}} 
%{\includegraphics[scale=0.5]{fig11b.eps}} 
{\includegraphics[scale=0.5]{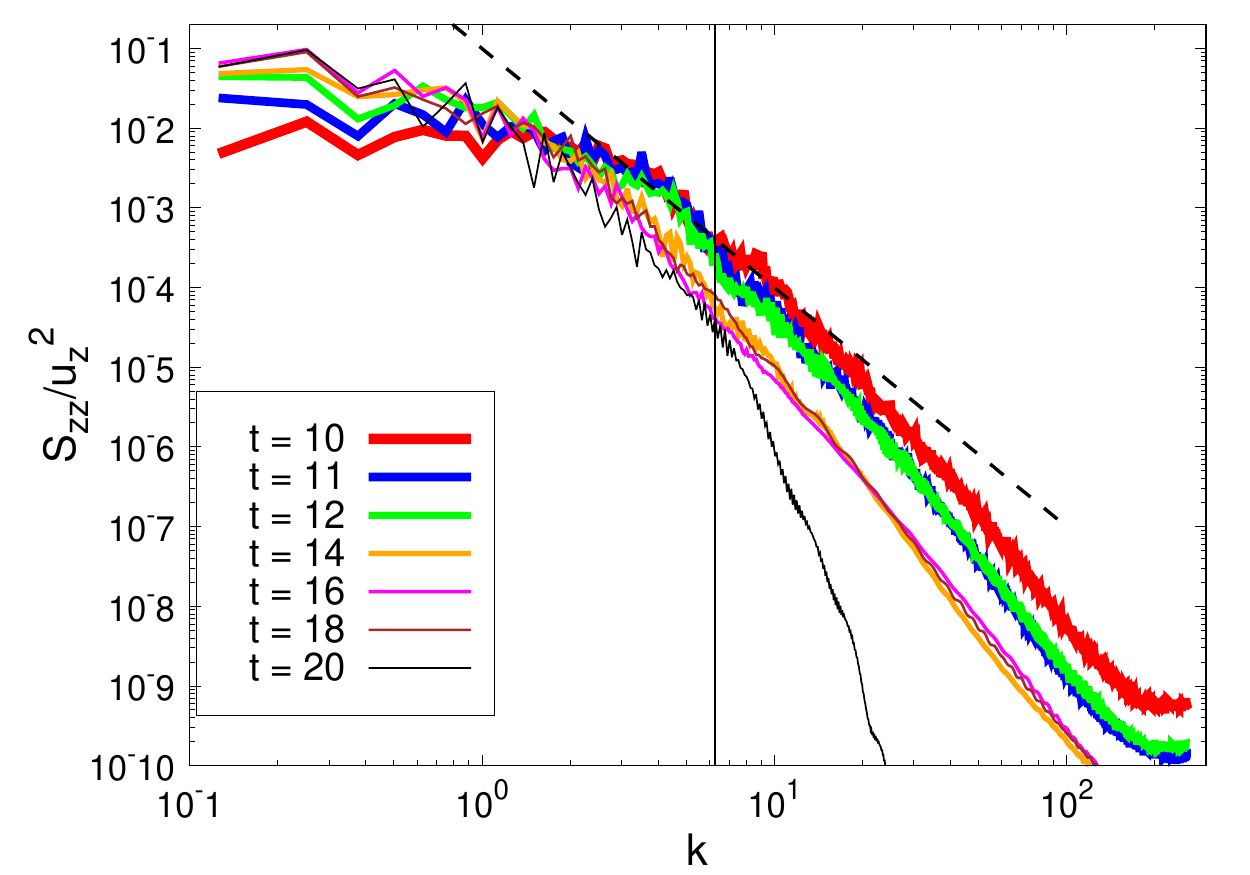}} 
%{\includegraphics[scale=0.5]{fig11d.eps}} 
\caption{Spectra of the vertical and horizontal component of the velocity for bubbles with 
(top) $Ar = 100$ and $Bo = 0.1$, 
%(bottom) $Ar = 140$ and $Bo = 0.2$; 
evaluated at different times in the interrogation window between $15$ and $20$ bubble diameters (in the vertical direction). The dashed line represents the $-3$ slope. The vertical line corresponds to the bubble diameter.}
\label{Fig:spettri-app}
\end{figure}

In figure \ref{Fig:spettri-app}, we show the same spectra as displayed in Fig. \ref{Fig:spettri2} but computed in the window between $15$ and $20$ bubble diameters.
It is clear that the same information content is present in both windows.
Moreover, thanks to the high resolution in time, the agitation is frozen for some time.
About five characteristic times after all bubbles have gone out from the interrogation window, the spectrum starts to decay exponentially.
%To show visually the dynamics corresponding to this time, we plot in Fig. \ref{Fig:vorticity-app} the vorticity at the last time $t=20$, that is when bubbles have long left and the flow is in the decaying regime.
In this regime the flow appears to be laminar with only some fluctuations at the largest scale. 
%\begin{figure}
%\center
%{\includegraphics[scale=0.725]{fig12c.eps}} {\includegraphics[scale=0.325]{colorbar.eps}}
%\caption{Vorticity field for bubbles with $Ar = 100$ and $Bo = 0.1$:
%At $t = 20$ the window is between $15$ and $20$ bubble diameters in the vertical direction.}
%\label{Fig:vorticity-app}
%\end{figure} 

\begin{figure}
%{\includegraphics[scale=0.5]{fig19a.eps}} 
%{\includegraphics[scale=0.5]{fig19b.eps}}
{\includegraphics[scale=0.5]{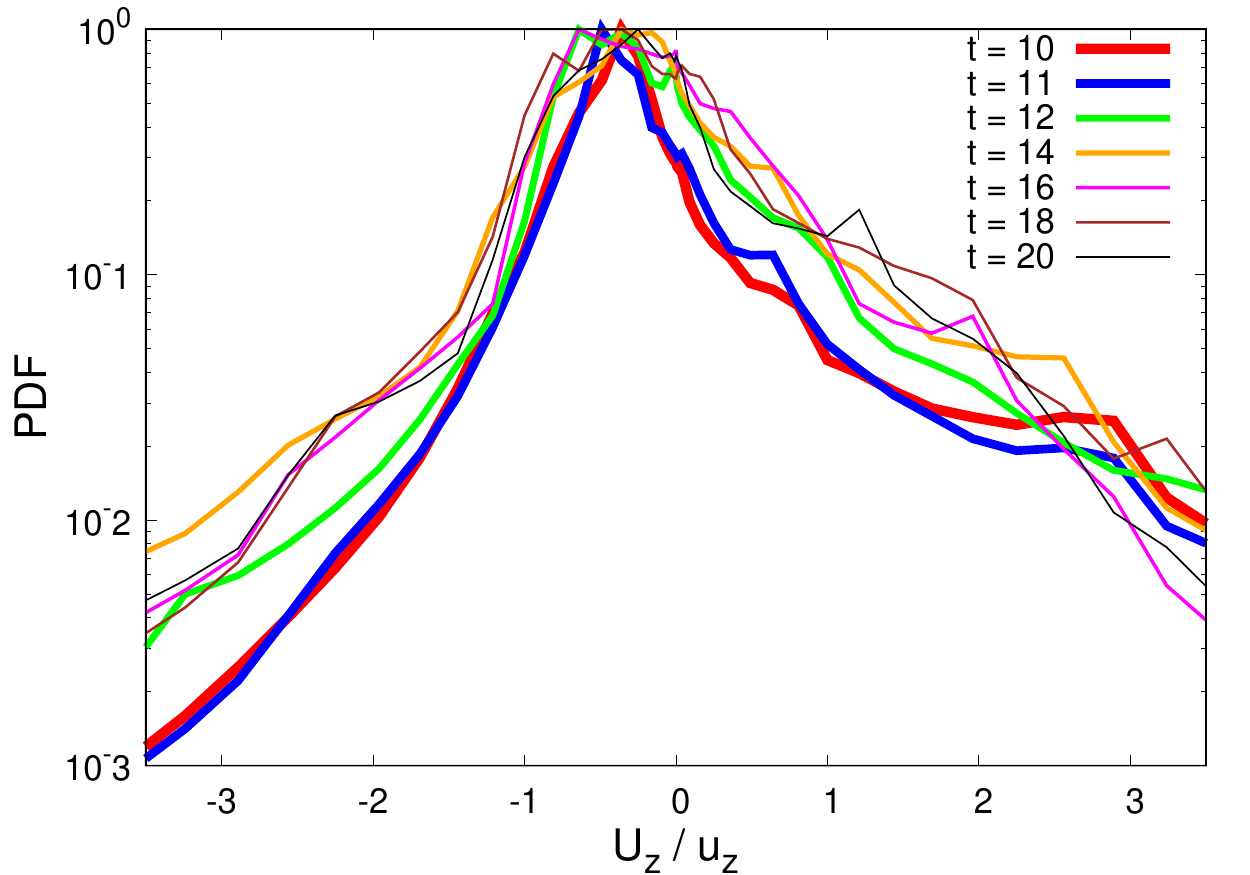}} 
{\includegraphics[scale=0.5]{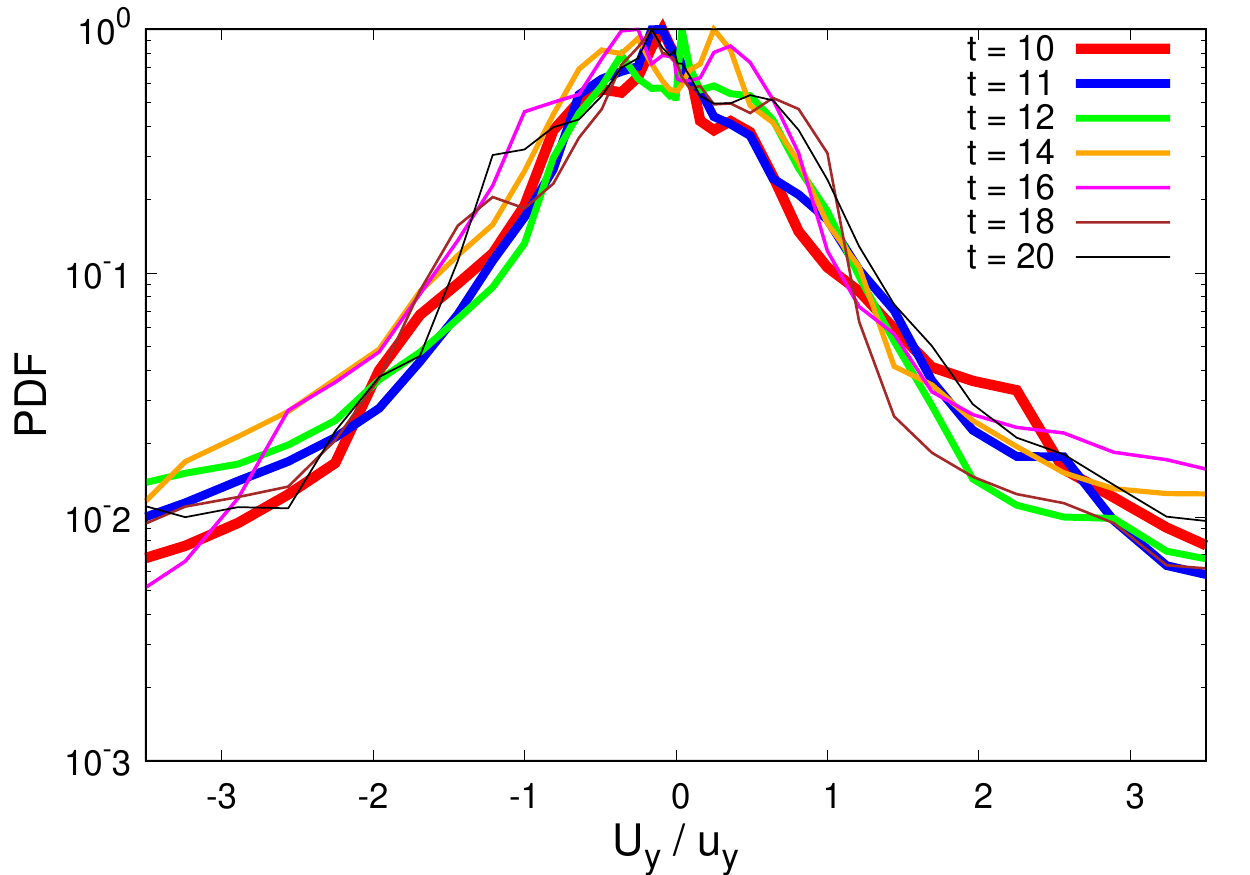}} 
\caption{PDF of the velocity fluctuations in the vertical $zz$ (left column) and lateral $yy$ (right column) directions. 
%Case (a) Bubbles at $Ar = 100$  and $Bo = 0.1$ (top panels); Case (b) 
Bubbles at $Ar = 140$  and $Bo = 0.2$ 
%(low panels)
. }
\label{Fig:pdf1}
\end{figure} 
%The $Ar=313$ case turns out to be turbulent. The bubbles are able to inject enough energy at the bubble diameter scale to trigger an inverse cascade of energy with its spectral $-5/3$ counterpart. 
%At smaller scales the situation is not well defined. The creation of smaller scales around the bubbles at some times is perceived as an injection of energy for the inverse cascade. At other times, the flux is basically zero at smaller scales and therefore a different spectrum may be found, notably the $-3$ spectrum typical of wakes and enstrophy production.
%However, in this range viscous damping is important and thus no neat scaling can be found except the viscous one.
%In order to get more insights about the cascade, we show in Figure \ref{Fig:filter1}  the spatial distribution of $\Pi_l$ (with $l = 0.5, 0.1 d_b$) for the case with $Ar=313$ at $t=12$. It can be noted how the regions of higher energy transfer are in the neighbourhood of the bubbles and how positive and negative areas are present over the whole domain. In particular smaller scales transfer is effective only in the vicinity of bubbles.
We now consider the fluctuation properties, 
looking at the probability density functions of the velocity fluctuations. 
Figures \ref{Fig:pdf1} show the PDFs of the two velocity components at various times for the case at $Ar=140$; results for the other $Ar$ numbers do not show appreciable differences.
While a time-dependent dynamics is noticeable due to the statistical unsteadiness of the flow, the results are basically the same starting from $t=13$, when most of the bubbles have transited through the spatial region where data are taken.

\end{document}